\documentclass[journal]{IEEEtran}
\usepackage[utf8]{inputenc}
\usepackage{graphicx}
\usepackage{amsmath}
\usepackage{amssymb}
\usepackage{mathtools}
\usepackage{accents}
\usepackage{verbatim}
\usepackage{enumitem,kantlipsum}
\usepackage{color}
\usepackage[noadjust]{cite}
\usepackage[caption=false,font=footnotesize]{subfig}
\newtheorem{prob}{Problem}
\newtheorem{lemma}{Lemma}
\newtheorem{definition}{Definition}
\newtheorem{proposition}{Proposition}

\newtheorem{theorem}{Theorem}
\newtheorem{assumption}{Assumption}

\ifCLASSINFOpdf
\else
\fi

\begin{document}

\title{Planning and Control of Multi-Robot-Object Systems  under Temporal Logic Tasks and Uncertain Dynamics}
\author{Christos K. Verginis, Yiannis Kantaros, and Dimos V. Dimarogonas
\thanks{C. K. Verginis is with the Division of Signals and Systems, Department of Electrical Engineering, Uppsala University, Uppsala, Sweden. Y. Kantaros is with the Department of Electrical and Systems Engineering, Washington University, St. Louis, MO, USA. D. V. Dimarogonas is with the Division of Decision and Control Systems, Department of Electrical Engineering and Computer Science, KTH Royal Institute of Technology, Stockholm, Sweden. }}% <-this %
%\date{May 2020}

\maketitle

\begin{abstract}
%\boldmath
We develop an algorithm for the motion and task planning of a system comprised of multiple robots and unactuated objects under tasks expressed as Linear Temporal Logic (LTL) constraints. The robots and objects evolve subject to uncertain dynamics in an obstacle-cluttered environment. The key part of the proposed solution is the intelligent construction of a coupled transition system that encodes the motion and tasks of the robots \textit{and} the objects. We achieve such a construction by designing appropriate adaptive control protocols in the lower level, which guarantee the safe robot navigation/object transportation in the environment while compensating for the dynamic uncertainties. The transition system is efficiently interfaced with the temporal logic specification via a sampling-based algorithm to output a discrete path as a sequence of synchronized actions of the robots; such actions satisfy the robots' as well as the objects' specifications. The robots execute this discrete path by using the derived low level control protocol.
%, allowing the interface with the task specification.    
%In particular, we design control protocols that allow the transition of the agents as well as the cooperative transportation of the objects by the agents, among predefined regions of interest in the workspace. This allows to abstract the coupled behavior of the agents and the objects as a finite transition system and to design a high-level multi-agent plan that satisfies the agents' and the objects' specifications, given as temporal logic formulas. 
Simulation results verify the proposed framework. 
\end{abstract}

\begin{IEEEkeywords}
Multi-agent systems, multi-agent-object systems, linear temporal logic, robot navigation, cooperative manipulation, object transportation, multi.
\end{IEEEkeywords}

% For peer review papers, you can put extra information on the cover
% page as needed:
% \ifCLASSOPTIONpeerreview
% \begin{center} \bfseries EDICS Category: 3-BBND \end{center}
% \fi
%
% For peerreview papers, this IEEEtran command inserts a page break and
% creates the second title. It will be ignored for other modes.
\IEEEpeerreviewmaketitle

\section{Introduction}
\IEEEPARstart{T}emporal-logic-based motion planning has gained significant amount of attention over the last decade, as it provides a fully automated  correct-by-design controller synthesis approach for autonomous robots. Temporal logics, such as linear temporal logic (LTL), provide formal high-level languages that can describe planning objectives more complex than the well-studied navigation algorithms, and have been used extensively both in single- as well as in multi-robot setups (see, indicatively, \cite{Fainekos2009,Lahijanian2016,Loizou2004,Diaz2015,Chen2012,Cowlagi2016,Belta2005,Bhatia2011,Filippidis2012,Meng15}). The task is given as a temporal logic formula, which is coupled with a discrete representation of the underlying system, abstracted from the underlying continuous dynamics, in order to derive an appropriate high-level discrete path. 

%and then, a high-level discrete path is found by optimization or model-checking algorithms, given the abstracted system and the task specification \cite{baier2008principles, kantaros2018sampling}.  

%is the derivation of a discrete transition system that couples the behavior of the robots and the objects in the workspace. We first combine point-world transformations and adaptive control
%to design a feedback control scheme that guarantee i) the navigation of the agents and ii) the cooperative transportation of the objects by the agents, among predefined regions of interest in the workspace, while ensuring collision avoidance with other robots and with the workspace obstacles. This implies the derivation of action and motion primitives, which we use to abstract the behavior of the robots and the objects by constructing a finite transition system. Finally, we use a sampling-based procedure to search the composition of the transition system and the task's automaton for the derivation of an optimal high-level plan, as a sequence of robot actions, that satisfies the task.

There exist numerous works that consider temporal-logic-based tasks, both for single- and multi-robot systems \cite{Fainekos2009,Lahijanian2016,Loizou2004,Chen2012,Belta2005,Filippidis2012}. 
Nevertheless, the related works consider temporal logic-based motion planning for fully actuated, autonomous robotic agents. Consider, however, cases where some unactuated objects must undergo a series of processes in a workspace with autonomous robots (e.g., car factories or automated warehouses). In such cases, the robots, except for satisfying their own motion specifications, are also responsible for coordinating with each other in order to transport the objects around the workspace. When the unactuated objects' specifications are expressed using temporal logics, 
then the abstraction of the robots' behavior becomes much more complex, since it has to take into account the objects' goals. 

In addition, the spatial discretization of a multi-agent system to an abstracted higher level system necessitates the design of appropriate continuous-time controllers for the transition of the agents among the states of discrete system. Many works in the related literature, however, either assume that there \textit{exist} such continuous controllers or adopt non-realistic assumptions. For instance, many works either do not take into account continuous agent dynamics or consider single or double integrators \cite{Loizou2004,Filippidis2012,Fainekos2009,Bhatia2011,Meng15}, which can deviate from the actual dynamics of the agents, leading thus to poor performance in real-life scenarios. 
Discretized abstractions, including design of the discrete state space and/or continuous-time controllers, can be found in \cite{Belta2005,belta2006controlling,reissig2011computing,tiwari2008abstractions,rungger2015state} for general systems and \cite{boskos2015decentralized,belta2004abstraction} for multi-agent systems. Moreover, many works adopt dimensionless point-mass agents and therefore do not consider inter-agent collision avoidance \cite{Belta2005,Filippidis2012,Meng15}, which can be a crucial safety issue in applications involving autonomous robots. 

Since we aim at incorporating the unactuated objects' specifications in our framework, the robots have to perform (cooperative) transportation of the objects around the workspace, while avoiding collisions with obstacles. Cooperative transportation/manipulation has been extensively studied in the literature (see, for instance, \cite{sugar2002control,heck2013internal,kume2007coordinated,tsiamis2015cooperative,ficuciello2014cartesian,ponce2016cooperative,marino2017distributed,nikou2017nonlinear,verginis2018communication,verginis2019robust,erhart2016model}), with collision avoidance specifications being incorporated in \cite{tanner2003nonholonomic,nikou2017nonlinear,verginis2018communication}. %, which is the main inspiration of our cooperative transportation methodology. 
Cooperative object transportation under temporal logics has also been considered in our previous work \cite{verginis2017distributed}. 

This paper proposes a novel algorithm for the motion and task planning of a multi-robot-object system, i.e., a system that consists of multiple robots and unactuated objects, subject to tasks that are expressed by LTL constraints. We consider that the robots and the objects evolve subject to continuous-time and -space dynamics that suffer from uncertainties and in an environment cluttered with obstacles. Our contribution is summarized as follows. 
First, we abstract the continuous dynamics of the multi-robot-object system into a finite transition system that encodes the coupled  behavior of the robots and the objects in the environment. We achieve such an abstraction by developing continuous feedback-control protocols that guarantee i) the navigation of the robots and ii) the cooperative transportation of the objects by the robots in the environment. These protocols combine barrier functions and point-world transformations to guarantee collision-avoidance properties as well as 
adaptive-control methodologies to compensate for the dynamic uncertainties. The robot navigation and cooperative object transportation constitute robot action primitives that enable the transitions in the aforementioned transition system.
Second, we compose the resulting transition system with an automaton-based representation of the underlying LTL task.
Finally, we use a sampling-based procedure to search the composition of the transition system and the task's automaton for the derivation of an optimal high-level plan, as a sequence of robot actions, that satisfies the task.

This paper extends our previous work \cite{verginis2017ifacPlann} in the following directions: firstly, we consider transportation of the objects by multiple robots (as opposed to  \cite{verginis2017ifacPlann}, which consider single-robot object transportation). Secondly, we consider more complex, obstacle-cluttered environments. Thirdly, we consider dynamic uncertainty in the dynamics of the robots and objects. Finally, instead of model-checking tools based on graph search algorithms, we use a sampling-based procedure to derive an optimal task-satisfying plan, yielding significantly lower complexity in terms of runtime and memory.

The rest of the paper is organized as follows. Section \ref{sec:Model and PF} provides necessary background and formulates the problem, while Section \ref{sec:main results} presents the proposed solution. Section \ref{sec:simulations} provides simulation results and, finally, Section \ref{sec:conclusion} concludes the paper.

\section{Preliminaries and Problem Formulation} \label{sec:Model and PF}
\subsection{Task Specification in LTL}
We focus on the task specification $\phi$ given as a Linear
Temporal Logic (LTL) formula. The basic ingredients of a LTL
formula are a set of atomic propositions $\Psi$ and several boolean and temporal operators. LTL formulas are formed according  to the following grammar \cite{baier2008principles}: $\phi :\coloneqq \mathrm{true} \ | \ a \ | \ \phi_1 \ \land \phi_2 \ | \ \neg \phi \ | \ \circ \phi \ | \ \phi_1 \mathcal{U} \phi_2$, where $a \in \Psi$, $\phi_1$ and $\phi_2$ are LTL formulas and $\circ$, $\mathcal{U}$ are the next and until operators, respectively. Definitions
of other useful operators like $\square$ (always), $\Diamond$ (eventually) and
$\Rightarrow$ (implication) are omitted and can be found at \cite{baier2008principles}. The
semantics of LTL are defined over infinite words over $2^\Psi$.
Intuitively, an atomic proposition $\psi \in \Psi$ is satisfied on a word
$w = w_1 w_2 \dots$ if it holds at its first position $w_1$, i.e. $\psi \in w_1$.
Formula $\phi$ holds true if $\phi$ is satisfied on the word suffix
that begins in the next position $w_2$, whereas $\phi_1 \mathcal{U} \phi_2$ states
that $\phi_1$ has to be true until $\phi_2$ becomes true. Finally, $\Diamond \phi$ and
$\square \phi$ holds on $w$ eventually and always, respectively. For a full
definition of the LTL semantics,  we refer the reader to  \cite{baier2008principles}.

\subsection{Problem Formulation} \label{subsec:pf}
Consider $N>1$ robotic agents operating in a compact $2$D workspace $\mathcal{W} \subset \mathbb{R}^2$ with an outer boundary with $M>0$ objects, and $\mathcal{N}\coloneqq \{1,\dots,N\}$, $\mathcal{M}\coloneqq \{1,\dots,M\}$; The objects are represented by rigid bodies whereas the robotic agents are fully actuated and holonomic, equipped with a transportation tool, such as a robotic arm.  In addition, the workspace is populated with $J\in\mathbb{N}$ connected, closed sets $\{ \mathfrak{O}_j \}_{j\in\mathcal{J}}$, with $\mathcal{J}\coloneqq \{1,\dots,J\}$, representing obstacles, and we define the free space as $\mathcal{W}_\textup{f} \coloneqq \mathcal{W} \backslash \bigcup_{j\in\mathcal{J}} \mathfrak{O}_j$.

We further assume that there exist $K>1$ points within $\mathcal{W}_\textup{free}$, denoted by $p_{\pi_k}$, corresponding to certain properties of interest (e.g., gas station, repairing area, etc.), with $\mathcal{K} \coloneqq \{1,\dots,K\}$. Since, in practise, these properties are naturally inherited to some neighborhood of the respective point of interest, we define for each $k\in\mathcal{K}$ the \textit{region of interest} $\pi_k$, corresponding to $p_{\pi_k}$, as the closed ball $\pi_k \coloneqq \bar{\mathcal{B}}(p_{\pi_k},r_{\pi_k}) \subset \mathcal{W}_\textup{free}$, with $r_{\pi_k} > 0$ positive radii, $\forall k\in\mathcal{K}$.
These properties of interest are expressed as boolean variables via finite sets of atomic propositions. 
In particular, we introduce disjoint sets of atomic propositions $\Psi_i, \Psi^{o}_j$, expressed as boolean variables, that represent services provided to agent $i\in\mathcal{N}$ and object $j\in\mathcal{M}$ in $\Pi \coloneqq \{\pi_1,\dots,\pi_K\}$.
The services provided at each region $\pi_k$ are given by the labeling functions $\mathcal{L}_i:\Pi\rightarrow2^{\Psi_i}, \mathcal{L}^{o}_j:\Pi\rightarrow2^{\Psi^{o}_j}$, which assign to each region $\pi_k, k\in\mathcal{K}$, the subset of services $\Psi_i$ and $\Psi^{o}_j$, respectively, that can be provided in that region to agent $i\in\mathcal{N}$ and object $j\in\mathcal{M}$, respectively. In addition, we consider that the agents and the object are initially ($t=0$) in the regions of interest $\pi_{init(i)}, \pi_{init_{o}(j)}$, where the functions $init:\mathcal{N}\to\mathcal{K}$, $init_{o}:\mathcal{M}\to\mathcal{K}$ specify the initial region indices. 
In the following, we present the modeling of the coupled dynamics of the object and the robots.
%\begin{figure}
%\centering
%\includegraphics[width = 0.25\textwidth]{robot_rigid_ellips.eps}
%
%\caption{ A robotic agent consisting of $\mathfrak{p}_i = 3$ rigid bodies, with $\mathfrak{R}^i_1 = 21$ and $\mathfrak{R}^i_2 = \mathfrak{R}^i_3 = 4$ ellipsoids. \label{fig:robot_rigid_ellips}}
%\end{figure}

%\subsection{Robotic Agents} \label{subsec:Robotics Agents}

We denote by $x_i \in \mathbb{R}^2$ the position of robot $i$'s center of mass. 
In this work we explicitly consider the actions of robot navigation, as well as (cooperative) transportation of an object via $x_i$, where the joint variables of the potential mounted robotic arm are assumed to be \textit{fixed}. We consider that the robotic arm joints are used only for grasping/releasing an object (when the respective mobile base is fixed), actions that we do not explicitly model. The dynamics describing the motion of robot $i$'s center of mass are given by 
\begin{align}
m_i\ddot{{x}}_i+f_i(x_i,\dot{{x}}_i) = u_i - h_i,\label{eq:agent dynamics navigation} 
\end{align}
where $m_i$ is the \textit{unknown} mass of robot $i$, $f_i:\mathbb{R}^{4}\to\mathbb{R}^2$ are \textit{unknown} friction-like functions, $u_i\in\mathbb{R}^2$ is the control input of robot $i$, and $h_i \in \mathbb{R}^2$ is the force exerted by robot $i$ at the grasping point with object $j$ in case of contact. The aforementioned dynamics concern the cases when (i) robot $i$ is navigating to some pre-defined point, and (ii) robot $i$ is transporting, possibly collaboratively with other robots, an object $j$. In both of these cases, the joint variables of the mounted robotic arm are assumed to be constant. The procedures of grasping/releasing an object, where the robotic arm would have to be activated, are not considered here. 

We consider that each robot $i$, for a given ${x}_i$, covers a spherical region %$\mathcal{A}_i\subset \mathbb{R}^2$ 
of constant radius $r_i\in\mathbb{R}_{>0}$ that bounds its volume, i.e., $\bar{\mathcal{B}}({x}_i,r_i)$, $\forall i\in\mathcal{N}$; 
Moreover, we consider that the agents have specific power capabilities, which for simplicity, we associate to positive integers $\zeta_i > 0$, $i\in\mathcal{N}$, via an analogous relation.
The overall configuration is %${q} \coloneqq [{q}^\top_1,\dots,{q}^\top _N]^\top$, 
${x}\coloneqq [{x}^\top_1,\dots,{x}^\top _N]^\top$ $\in\mathbb{R}^{2N}$.

Regarding the objects, we denote by $x^o_j \in \mathbb{R}^2$ the position of the $j$th object's center of mass, $\forall j\in\mathcal{M}$. We consider the following second-order Newton-Euler dynamics: 
%\begin{subequations} 
\begin{align} \label{eq:object dynamics}
m^{o}_j\ddot{{x}}^{o}_j + f^o_j(x^o_j,\dot{x}^{o}_j) = h^{o}_j,
\end{align}
%\end{subequations}
where $m^o_j$ is the \textit{unknown} mass of object $j$, $f^o_j : \mathbb{R}^4 \to \mathbb{R}^2$ are \textit{unknown} friction terms, and $h^o_j \in \mathbb{R}^2$ are the forces exerted to the $j$th object's center of mass, in case of contact with a robot. The overall object configuration is denoted by $x^o \coloneqq [x^o_1,\dots,x^o_M]^\top$, and $\bar{x} \coloneqq [x^\top, (x^o)^\top]^\top$.

The functions $f_i$ and $f^o_j$ are assumed to satisfy the following assumption:
\begin{assumption} \label{ass:friction}
The functions $f_i,f^o_j:\mathbb{R}^{4}\to\mathbb{R}^2$ are analytic and satisfy 
\begin{align} \label{eq:f_i ass}
    &\|f_i(\mathsf{x},\mathsf{y})\| \leq \alpha_i \|\mathsf{y}\|,\\
    &\|f^o_j(\mathsf{x},\mathsf{y})\| \leq \alpha^o_j \|\mathsf{y}\|,
\end{align}
$\forall \mathsf{x},\mathsf{y} \in \mathbb{R}^2$, where $\alpha_i$, $\alpha^o_j$ are unknown positive constants, $\forall i\in\mathcal{N},j\in\mathcal{M}$.
\end{assumption}
The aforementioned assumption is inspired by standard friction-like terms, which can be approximated by continuously differentiable velocity functions \cite{makkar2005new}. 

%${M}_{o}:\mathbb{M}\to\mathbb{R}^{6\times6}$ is the positive definite inertia matrix, ${C}_{o}:\mathbb{M}\times\mathbb{R}^{6}\to\mathbb{R}^{6\times6}$ is the Coriolis matrix, ${g}_{o}:\mathbb{M}\to\mathbb{R}^6$ is the gravity vector, and ${f}^{o}_j\in\mathbb{R}^6$ is the vector of generalized forces in case of contact with the external environment, $\forall j\in\mathcal{M}$; $S(\cdot$ is the skew-symmetric operator defined according to the cross product $S(a)b = a \times b$ for any $a, b\in\mathbb{R}^3$. 

Similarly to the robots, each object's volume is represented by the spherical set %$\mathcal{O}_j \subset \mathbb{R}^2$ 
of constant radius $r^{o}_j \in\mathbb{R}_{>0}$, i.e., %$\mathcal{O}_j({x}^{o}_j)  \coloneqq$ 
$\bar{\mathcal{B}}({x}^{o}_j,r^{o}_j)$, $\forall j\in\mathcal{M}$. 

Next, we provide the coupled dynamics between an object $j\in\mathcal{M}$ and a subset $\mathcal{A}\subseteq\mathcal{N}$ of robots that grasp it rigidly.
For these robots, it holds that $h^o_j = \sum_{i\in\mathcal{A}} h_i$, and
since the joint variables of the robotic arms are fixed,  $x^o_j = x_i + d^o_{ij}$, $\dot{x}^o_j = \dot{x}_i$ and $\ddot{x}^o_j = \ddot{x}_i$, where $d^o_{ij} \in \mathbb{R}^2$ is the constant offset between $x^o_j$ and $x_i$, $\forall i\in\mathcal{A}$.
Therefore, one obtains that
\begin{equation}\label{eq:coupled dynamics}
    {m}_{\mathcal{A},j} \ddot{x}^o_j + {f}_{\mathcal{A},j}(x^o_j,\dot{x}^o_j) = \sum_{i\in\mathcal{A}} u_i,
\end{equation}
where ${m}_{\mathcal{A},j} \coloneqq m^o_j + \sum_{i\in\mathcal{A},j} m_i$, ${f}_{\mathcal{A},j} \coloneqq f^o_j(x^o_j, \dot{x}^o_j) + \sum_{i\in\mathcal{A},j} f_i(x^o_j - d^o_{ij}, \dot{x}^o_j)$. Note that Assumption \ref{ass:friction} implies that 
\begin{equation}
    \| {f}_{\mathcal{A},j}(x^o_j,\dot{x}^o_j) \| \leq \alpha_{\mathcal{A},j} \|\dot{x}^o_j\|,
\end{equation}
for an unknown positive constant $\alpha_{\mathcal{A},j}$.

Regarding the volume of the coupled robots-object system, we consider that it is bounded by a sphere of radius 
%denote by $\mathcal{AO}_{\mathcal{T},j} \subset \mathbb{R}^2$ the sphere 
centered at ${x}^{o}_j$ with constant radius $r_{\scriptscriptstyle \mathcal{A},j} \in\mathbb{R}_{>0}$, i.e., %$\mathcal{AO}_{\mathcal{T},j}({x}^{o}_j) \coloneqq 
$\mathcal{B}({x}^{o}_j,r_{\scriptscriptstyle \mathcal{A},j})$, which is large enough to cover the volume of the coupled system.  

Moreover, in order to take into account the introduced robots' power capabilities $\zeta_i$, $i\in\mathcal{N}$, we consider a function $\Lambda\in\{\mathsf{True},\mathsf{False}\}$ that outputs whether the robots that grasp an object are able to transport the object, based on their power capabilities. For instance, $\Lambda(m^{o}_j, \zeta_{\mathcal{A}}) = \mathsf{True}$, where $m^{o}_j \in\mathbb{R}_{>0}$ is the mass of object $j$ and $\zeta_{\mathcal{A}} \coloneqq [\zeta_i]^\top_{i\in\mathcal{A}}$, implies that the robots $\mathcal{A}$ have sufficient power capabilities to cooperatively transport object $j$. 
%Note that, in real-time scenarios, the function $\Lambda$ can be easily defined explicitly in view   
%\subsection{Problem Formulation} \label{subsec:PF}
%We formally describe now the problem tackled in this paper. We first introduce some preliminary required notation. 

Next, we define the boolean functions $\mathcal{AG}_j:\mathbb{R}_{\geq 0} \to 2^{\mathcal{N}_0}$, with $\mathcal{N}_0 \coloneqq \mathcal{N} \cup \{0\}$, to denote which robots grasp an object $j\in\mathcal{M}$; $\mathcal{AG}_j = 0$ means that no robots grasp object $j$.  %We also define $\mathcal{AG}_{i,0}\in\{\mathsf{},\mathsf{False}\}$, to denote that agent $i$ does not grasp any objects, i.e., $\mathcal{AG}_{i,j} =  \mathsf{False}, \forall j\in\mathcal{M} \Leftrightarrow \mathcal{AG}_{i,0} = \mathsf{}$, $\forall i\in\mathcal{N}$.
Note also that $i \in \mathcal{AG}_j \Leftrightarrow i \notin \mathcal{AG}_{j'}=\mathsf{False}, \forall j' \in \mathcal{M}\backslash\{j\}$, i.e., robot $i$ can grasp at most one object at a time. We further denote $\mathcal{AG} \coloneqq [\mathcal{AG}_1,\dots,\mathcal{AG}_M]^\top \in  (2^{\mathcal{N}_0})^M$.

In the following, we use the term ``entity" to refer to single robots, objects as well as systems comprised by robots that grasp an object (robots-object systems). The number of these systems depends on the variables $\mathcal{AG}$.
Given a grasping configuration $\mathcal{AG} \in  (2^{\mathcal{N}_0})^M$, consider $\bar{T}(\mathcal{AG})$ number of entities, indexed by the set $\mathcal{T}(\mathcal{AG}) \coloneqq \{1,\dots,\bar{T}(\mathcal{AG})\}$. Each entity (robot, object, or robots-object system) is characterized by the respective configuration ($x_i$, $x_j^o$, $x_{j'}^o$) and radius ($r_i$, $r^o_j$, $r_{\mathcal{A},j'}$), respectively, which we denote for simplicity by the generic variables $x_i^e$, $r_i^e$, for all $i\in\mathcal{T}(\mathcal{AG})$. 

We further define the free space for each entity $\mathcal{F}_i(\mathcal{AG}) \coloneqq \big( \mathcal{W}_\textup{f} \backslash \bigcup_{j\in\mathcal{T}(\mathcal{AG}) \backslash\{ i \} } \bar{\mathcal{B}}(x_j^e,r^e_j) \big) \ominus r^e_i
$, where the incorporation of $\ominus r^e_i$ enlarges the obstacles and the other entities with the radius $r_i^e$.

We give now
the definitions for the transitions of the robots and the objects between the regions of interest.

\begin{definition}(\textbf{Navigation})  \label{def:agent transition}
	Let $\mathcal{AG}(t_0) \in  (2^{\mathcal{N}_0})^M$ such that 
	\begin{align*}
	\left.
	\begin{matrix*}[l]
	& x^e_i(t_0) \in \mathcal{F}_i(\mathcal{\mathcal{AG}}(t_0))    \\
	& \bar{\mathcal{B}}_i(x_i^e(t_0),r_i^e) \subset \pi_{i_k}
	\end{matrix*} \ \ \ \right\} i\in\mathcal{T}(\mathcal{AG}(t_0))
	\end{align*}
	where $i_k \in \mathcal{K}$, for all $i\in\mathcal{T}(\mathcal{AG})$ and some $t_0\in\mathbb{R}_{\geq 0}$. Then, entity $j\in \mathcal{T}(\mathcal{AG})$ executes a transition from $\pi_{j_k}$ to $\pi_{j_{k'}}$, with $j_{k'} \in \mathcal{K} \backslash \{j_k\}$, if there exists a finite $t_f\geq t_0$
	such that 
	\begin{align*}
	& \bar{\mathcal{B}}_j(x_j^e(t_f),r_j^e) \subset \pi_{j_{k'}}    \\
	& x_j(t) \in \mathcal{F}_j(\mathcal{AG})
	\end{align*}
	for all $t\in[t_0,t_f]$.
	%\begin{equation}
	%\mathcal{A}_i({q}_i(t))\cap \Bigg(\Big(\bigcup\limits_{n\in\mathcal{N}\backslash\{i\}}\mathcal{A}_n({q}_n(t))\Big)\cup \Big(  \bigcup\limits_{j\in\mathcal{M}} \mathcal{O}_j({x}_j(t)) \Big)\cup \Big( \bigcup\limits_{m\in\mathcal{K}_\mathcal{R}\backslash\{k,k'\}} \pi_m \Big)\Bigg) = \emptyset,
	%\end{equation}	
	%$ \forall t\in[t_0, t_f]$.
\end{definition}

\begin{definition}(\textbf{Grasping}) \label{def:grasping}
   Let $\mathcal{AG}(t_0) \in  (2^{\mathcal{N}_0})^M$ such that 
	\begin{align*}
	&\left. \begin{matrix*}[l]
	& x^e_i(t_0) \in \mathcal{F}_i(\mathcal{AG}(t_0))    \\
	& \bar{\mathcal{B}}_i(x_i^e(t_0),r_i^e) \subset \pi_{i_k} 
	\end{matrix*} \ \ \ \right\} i\in\mathcal{T}(\mathcal{AG}(t_0)) \\
    &\hspace{5mm} j \notin \mathcal{AG}_\ell(t_0) \\
	&\hspace{5mm} x_{j}(t_0), x_\ell^o(t_0) \in \pi_{\ell}
	\end{align*}
	for some $j\in\mathcal{N}$, $\ell\in\mathcal{M}$, $\ell \in \mathcal{K} $, $t_0\in\mathbb{R}_{\geq 0}$
	where $i_k \in \mathcal{K}$, for all $i\in\mathcal{T}(\mathcal{AG})$. 
	Then, agent $j$ \textit{grasps} object $\ell$, denoted by $j \xrightarrow{g} \ell$, if there exists a finite $t_f\geq t_0$ such that $j \in \mathcal{AG}_{\ell}(t_f)$.
\end{definition}
Similarly, we can define the \textit{releasing} action $j \xrightarrow[]{r} \ell$ for an agent $j \in \mathcal{N}$ and object $\ell\in\mathcal{M}$.
Loosely speaking, the aforementioned definitions correspond to specific action primitives of the robots, namely robot navigation or object transportation, which define the \textit{navigation} transition, and grasping and releasing actions. 
%which define the respective transitions. 
When the navigation transition from $\pi_{j_k}$ to $\pi_{j'_k}$ corresponds to a robot navigation for robot $j\in\mathcal{N}$, we denote it by $\pi_{j_k} \rightarrow_j \pi_{j'_k}$; when it corresponds to a cooperative transportation of object $j\in\mathcal{M}$ by a subset of robots  $\mathcal{A}$, we denote it by $\pi_{j_k} \xrightarrow{T}_{\mathcal{A},j} \pi_{j'_k}$. 

We also assume the  existence of a procedure ${\mathcal{P}}_s$ that outputs whether or not a set of non-intersecting spheres fits in a larger sphere as well as possible positions of the spheres in the case they fit. More specifically, given a region of interest $\pi_k$ and a number $\widetilde{N}\in\mathbb{N}$ of sphere radii (of robots, objects, or object-robots systems) the procedure can be seen as a function ${\mathcal{P}}_s \coloneqq [\mathcal{P}_{s,0}, {\mathcal{P}}^\top_{s,1}]^\top$, where $\mathcal{P}_{s,0}:\mathbb{R}^{\widetilde{N}+1}_{\geq 0}\to\{\mathsf{True},\mathsf{False}\}$ outputs whether the spheres fit in the region $\pi_k$ whereas  ${\mathcal{P}}_{s,1}$ provides possible configurations of the robots and the objects 
or ${0}$ in case the spheres do not fit.
For instance, $P_{s,0}(r_{\pi_2},r_1,r_3,r^{o}_1,r^{o}_5)$ determines whether the robots $1,3$ and the objects $1,5$ fit in region $\pi_2$, without colliding with each other; $(x^e_1,x^e_2,x^e_3,x^e_4) = (x_1,x_3,{x}^{o}_1, {x}^{o}_5) =  P_{s,1}(r_{\pi_2},r^e_1,r^e_2,r^e_3,r^e_4)
=P_{s,1}(r_{\pi_2},r_1,r_3,r^{o}_1,r^{o}_5)$ provides a set of configurations such that $\bar{\mathcal{B}}_i(x^e_i,r^e_i)$ $i\in\{1,\dots,4\}$, and the pairwise intersections of these sets are empty.
The problem of finding an algorithm $\mathcal{P}_s$ is a special case of the sphere packing problem \cite{chen2008sphere}. Note, however, that we are not interested in finding the maximum number of spheres that can be packed in a larger sphere but, rather, in the simpler problem of determining whether a set of spheres can be packed in a larger sphere. 

%\subsection{Specification} \label{subsec:Specification}
Our goal is to control the multi-robot-object system defined above such that the robots and the objects obey a given specification over their atomic propositions $\Psi_i, \Psi^{o}_j, \forall i\in\mathcal{N},j\in\mathcal{M}$. 
%We define  $\bar{{p}}:\mathbb{R}^{\mathfrak{n}}\times\mathbb{R}_{\geq 0}\rightarrow\mathbb{R}^{N+M}$ with $\bar{{p}}({q},t)=[{p_1}({q_1}),\dots,{p_1}({q_N}), {p_{o_1}}(t), \dots, {p_{o_M}}(t)]^T$, and ${q} = [{q^T_1},\dots,{q^T_N}]^T\in\mathbb{R}^{\mathfrak{n}}, \mathfrak{n} = \sum\limits_{i\in\mathcal{N}} \mathfrak{n}_i$.
Given the trajectories ${x}_i(t), {x}^{o}_j(t), t\in\mathbb{R}_{\geq 0}$, of robot $i$ and object $j$, respectively, their corresponding \textit{behaviors} are given by the infinite sequences 
\begin{align*}
& b_i \coloneqq ({x}_i(t),\sigma_i) \coloneqq ({x}_i(t_{i,1}),\sigma_{i,1})({x}_i(t_{i,2}),\sigma_{i,2})\dots, \\ 
& b^{o}_j \coloneqq ({x}^{o}_j(t),\sigma^{o}_j) \coloneqq ({x}^{o}_j(t^{o}_{j,1}),\sigma^{o}_{j,1}) ({x}^{o}_j(t^{o}_{j,2}),\sigma^{o}_{j,2})\dots,
\end{align*}
 with $t_{i,\ell+1} > t_{i,\ell} \geq 0, t^{o}_{j,\ell+1} > t^{o}_{j,\ell} \geq 0, \forall \ell\in\mathbb{N}$, representing specific time stamps. The sequences $\sigma_i, \sigma^{o}_j$ are the services provided to the agent and the object, respectively, over their trajectories, i.e., $\sigma_{i,\ell}\in 2^{\Psi_i}, \sigma^{o}_{j,l}\in 2^{\Psi^{o}_j}$ with $\mathcal{A}_i({q}_i(t_{i,\ell})) \subset \pi_{k_{i,\ell}}, \sigma_{i,\ell} \in \mathcal{L}_i(\pi_{k_{i,\ell}})$ and $\mathcal{O}_j({x}^{o}_j(t^{o}_{j,l})) \subset \pi_{k^{o}_{j,l}}, \sigma^{o}_{j,l} \in \mathcal{L}^{o}_j(\pi_{k^{o}_{j,l}}), k_{i,\ell}, k^{o}_{j,l} \in\mathcal{K}, \forall \ell,l\in\mathbb{N}, i\in\mathcal{N}, j\in\mathcal{M}$, where $\mathcal{L}_i$ and $\mathcal{L}^{o}_j$ are the previously defined labeling functions. The following Lemma then follows: 
\begin{lemma}
The behaviors $b_i, b^{o}_j$ satisfy formulas $\phi_i, \phi_{o_j}$ if $\sigma_i \models \phi_i$ and $\sigma^{o}_j \models \phi^{o}_j$, respectively.
\end{lemma}

%\subsection{Problem Formulation}
The control objectives are given as LTL formulas $\phi_i, \phi^{o}_j$ over $\Psi_i, \Psi^{o}_j$, respectively, $\forall i\in\mathcal{N},j\in\mathcal{M}$. The LTL formulas $\phi_i, \phi^{o}_j$ are satisfied if there exist behaviors $b_i,b^{o}_j$ of agent $i$ and object $j$ that satisfy  $\phi_i, \phi^{o}_j$. We are now ready to give a formal problem statement:
\begin{prob} \label{problem}
Consider $N$ robotic agents and $M$ objects in $\mathcal{W}$ satisfying 
 \begin{align*}
  & \bar{\mathcal{B}}_i(x_i(0),r_i)\subset \pi_{\text{init}(i)}, \bar{\mathcal{B}}_j({x}^{o}_j(0),r^{o}_j) \subset\pi_{\text{init}_{o}(j)}, \dot{{x}}_i(0) = {0},\\ 
  & \dot{x}^{o}_{j} = {0}, \forall i\in\mathcal{N},j\in\mathcal{M},
   \end{align*}
in collision-free initial configurations.
%where $\text{init}:\mathcal{N}\to\mathcal{K}$, $\text{init}_{o}:\mathcal{M}\to\mathcal{K}$ are functions providing the initial regions of interest, 
%\item $\mathcal{CNote that it is implicit in the problem statement the fact that the agents/objects starting in the same region can actually fit without colliding with each other. Technically, it holds that $\mathcal{P}_{s,0}(r_{\pi_k},[r_i]_{i\in \{i\in\mathcal{N}: \text{init}(i)= k\} }, [r^{o}_j]_{j\in \{ j\in\mathcal{M}: \text{init}_{o}(j) = k\} }) = \mathsf{True}$, $\forall k\in\mathcal{K}$.
%}_{i,l}({q}_i(0),{q}_l(0)) = \mathcal{C}_{o_j,o_\ell}({x}^{o}_j(0),{x}^{o}_\ell(0)) = \mathcal{C}_{i,o_j}({q}_{i}(0),{x}^{o}_\ell(0)) =  \mathsf{False}$, $\forall i,l\in\mathcal{N}, i\neq l, j,\ell\in\mathcal{M},j\neq \ell$.  
%\end{enumerate}
Given the disjoint sets $\Psi_i,\Psi^{o}_j$, $N$ LTL formulas $\phi_i$ over $\Psi_i$ and $M$ LTL formulas $\phi^{o}_j$ over $\Psi^{o}_j$, develop a control strategy that achieves behaviors $b_i, b^{o}_j$ which yield the satisfaction of 
$\phi_i, \phi^{o}_j, \forall i\in\mathcal{N},j\in\mathcal{M}$.
\end{prob}

Note that it is implicit in the problem statement the fact that the agents/objects starting in the same region can actually fit without colliding with each other. Technically, it holds that $\mathcal{P}_{s,0}(r_{\pi_k},[r_i]_{i\in \{i\in\mathcal{N}: \text{init}(i)= k\} }, [r^{o}_j]_{j\in \{ j\in\mathcal{M}: \text{init}_{o}(j) = k\} }) = \mathsf{True}$, $\forall k\in\mathcal{K}$.

\section{Main Results} \label{sec:main results}

\subsection{Continuous Control Design}
The first ingredient of our solution is the development of feedback control laws that establish the navigation transition of Def. \ref{def:agent transition}, i.e., robot navigation and cooperative object transportations. We do not focus on the grasping/releasing actions (see Def. \ref{def:grasping}) and we refer to some existing methodologies that can derive the corresponding control laws (e.g., \cite{Cutkosky2012,Reis2015}). 
%In the following, we use the term ``entities" to refer to the robots or the object-robot systems (multiple robots that grasp an object). 

The control design is based on the integration of the adaptive control scheme and point world transformation proposed in \cite{verginis2021adaptive} and \cite{vlantis2018robot}, respectively. The former accommodates the uncertain robot and object dynamics and the latter deals with the complex environment obstacles. 
Since \cite{vlantis2018robot} consider single-robot navigation in complex environments, we consider here a \textit{sequential} execution of the navigation and object-transportation transitions\footnote{The correctness of the multi-robot adaptive control scheme of \cite{verginis2021adaptive} has only been proven for {simple, spherical} environments.}. That is, only one entity is allowed to navigate from one RoI to another, viewing the rest of the entities as fixed obstacles.  \\

\noindent \textbf{{1) Robot Navigation}}

Assume that the conditions of Problem \ref{problem} hold for some $t_0 \in\mathbb{R}_{\geq 0}$, i.e., all robots and objects are located in some RoI with zero velocity.
We first consider the control problem of safe navigation for a robot $i\in\mathcal{N}$ satisfying $i \notin \mathcal{AG}_{j}(t_0)$ for all $j\in\mathcal{M}$, i.e., not grasping any objects, from region $i_k$ to $i'_k$ (transition $\pi_{i_k} \rightarrow_i \pi_{i'_k}$), with $\bar{\mathcal{B}}_i(x_i(t_0),r_i) \subset \pi_{i_k}$ and $i_k$, $i'_k$
$\in \mathcal{K}$. 
%Consider first an agent $i$, subject to the dynamics \eqref{eq:agent dynamics navigation}, aiming to safely navigate to a point $x_\text{d}$ of a neighboring cell. 
Let $x^e_i = x_i$. As stated before, the other entities are fixed and viewed as static obstacles by robot $i$. Moreover, to account for safety specifications, we wish the robot to avoid entering the RoI $\pi_k$, $k\in\mathcal{K}\backslash\{i_k,i'_k\}$. Therefore, the free space of robot $i$ becomes $\mathcal{F}_i = (\mathcal{W}_\textup{f} \backslash \mathcal{O}_i )\ominus r_i$, where
\begin{align*}
  \mathcal{O}_i \coloneqq  \left(\bigcup_{\ell\in\mathcal{T}(\mathcal{AG})\backslash\{i\}} \bar{\mathcal{B}}(x^e_\ell,r^e_\ell) \right)  \bigcup \left( \bigcup_{k\in\mathcal{K}\backslash\{i_k,i_k'\}} \pi_k  \right)  
\end{align*}
%Let $\{ \bar{\mathfrak{O}}_j \}_{j\in \mathcal{J}}$ be the set of the $J$ environment obstacles enlarged by the radius of the agent $r_i$, so that the latter reduces to the point $x_i$. As stated before, the other entities     are fixed and viewed as static obstacles by robot $i$. 
Let the desired navigation goal of the robot be $x_\textup{d,i} \in \pi_{i'_k}$, which will be provided by the procedure $\mathcal{P}_s$, as explained in the previous section. 
Next, we use the workspace transformation $\chi = \mathcal{H}(x)$ from \cite{vlantis2018robot} that converts the environment to the open unit disk $\mathcal{D} \coloneqq \mathcal{B}([0,0]^\top,1)$, and the obstacles (environment obstacles,  entities other than robot $i$, and RoI other than $\pi_{i_k}$, $\pi_{i'_k}$) to $\bar{J}$ points $b_\ell \in \mathcal{D}$, $\ell\in\bar{\mathcal{J}}\coloneqq \{1,\dots,\bar{J}\}$, robot $i$ to the point $\chi_i \coloneqq \mathcal{H}(x_i)$,
and the goal of robot $i$ to $\chi_\textup{d} \coloneqq \mathcal{H}(x_\textup{d,i})$. In the subsequent analysis we omit the robot index $i$ for ease of presentation.
%Hence, the free space of robot $i$ becomes $\mathcal{F}_i =  $
%Since there is not interaction with the other agents at this stage, we omit subscript $i$ in what follows.
%\red{Talk about the transformation $\chi = \mathcal{H}(x)$ that reduces the workspace to the open unit disk $\mathcal{D}$, the obstacles to $J$ points $b_j \in \mathcal{D}$, $j\in\mathcal{J}\coloneqq \{1,\dots,J\}$, and $x_\text{d}$ to $\chi_\text{d} \coloneqq \mathcal{H}(x_\text{d}) \in \mathcal{D}$ - if the desired point is on the boundary the adaptive controller needs modification.}

Let now %Note that, since the obstacles are disjoint, $b_j$, there exists 
a constant $\bar{r} > 0$ satisfying
\begin{align*}
    \|b_i - b_j \| > 2\bar{r}, \forall i,j \in\bar{\mathcal{J}}, i\neq j \\
    1 - \|b_i\| > 2\bar{r}, \forall i \in\bar{\mathcal{J}}.
\end{align*}
The transformed free space of the robot can be defined as $\mathcal{F}_\mathcal{H} \coloneqq \mathcal{D} \backslash\{b_1,\dots,b_{\bar{J}}\}$.
We define next  the set $\bar{\mathcal{J}}_0 \coloneqq \{0\}\cup\bar{\mathcal{J}}$ as well as the distances $d_\ell:\mathcal{F}_\mathcal{H} \to \mathbb{R}_{\geq 0}$, $\ell\in\bar{\mathcal{J}}_0$, with $d_\ell(\chi) \coloneqq \| \chi - b_\ell\|^2$, $\forall \ell\in\bar{\mathcal{J}}$, and $d_0(x) \coloneqq 1 - \| \chi \|^2$. Note that, by keeping $d_\ell(\chi) > 0$, $d_0(\chi) > 0$, we guarantee that $\chi \in \mathcal{F}_{\mathcal{H}}$ and hence the safety of the robot. 
%that $x_i \notin \mathcal{O}$, implying the safety of agent $i$.
	{We {also define} the constant 
	%\vspace{-6mm}	
	\begin{equation} \label{eq:bar_r_d}
%	\bar{r}_{\textup{d}} \coloneqq \min \left\{\bar{r}_{\mathcal{W}}^2-\|x_{\textup{d}}\|^2, \min_{j\in\mathcal{J}}\left\{ \|x_{\textup{d}} - c_j\|^2 - \bar{r}^2_{o_j} \right\} \right\}	
	\bar{r}_{\textup{d}} \coloneqq \min \left\{1- \|\chi_\textup{d}\|^2, \min_{\ell\in\bar{\mathcal{J}}}\left\{ d_\ell(\chi_\textup{d}) \right\} \right\}
	\end{equation}
	%\normalsize
	as the minimum distance of the goal to the transformed obstacles/workspace boundary.}
	We revisit now  the notion of the \textit{$2$nd-order navigation function} from \cite{verginis2021adaptive}: 
	\begin{definition}\label{def:2nd nf}
	A \textit{$2$nd-order navigation function} is a function $\phi: \mathcal{F}_\mathcal{H} \to \mathbb{R}_{\geq 0}$ of the form 

		\begin{equation*} %\label{eq:potential function}
			\varphi(\chi) \coloneqq k_1\|\chi - \chi_{\textup{d}}\|^2 + k_2\sum_{\ell\in\bar{\mathcal{J}}_0}\beta(d_\ell(\chi)),
		\end{equation*}		
		where $\beta:\mathbb{R}_{>0}\to\mathbb{R}_{\geq 0}$ is a (at least) twice contin. differentiable function and $k_1, k_2$ are positive constants, with the followings properties:
		\begin{enumerate}
			\item $\beta((0,\tau])$ is strictly decreasing,   				
			$\lim_{{\mathsf{x}} \to 0} {\beta}({\mathsf{x}}) = \infty$, and $\beta({\mathsf{x}}) = \beta(\tau)$, $\forall {\mathsf{x}} \geq \tau$,  for some $\tau > 0$, 
			\item $\varphi(\chi)$ has a global minimum  at $\chi = \chi_{\textup{d}} \in \textup{int}(\mathcal{F}_\mathcal{H})$ where $\varphi(
			\chi_{\textup{d}}) = 0$,			
			\item if $\beta'(d_k(\chi)) \neq 0$ {and $\beta''(d_k(\chi)) \neq 0$} for some $k\in\bar{\mathcal{J}}_0$, then $\beta'(d_\ell(\chi)) = \beta''(d_\ell(\chi)) = 0$, for all $\ell\in\bar{\mathcal{J}}_0 \backslash\{k\}$.
			%\item $\phi(x)$ has non-degenerate critical points, 
			\item The function $\widetilde{\beta}:(0,\tau) \to \mathbb{R}_{\geq 0}$, with
			%\begin{equation}
				$\widetilde{\beta}(\mathsf{x}) \coloneqq \beta''(\mathsf{x}) \mathsf{x}\sqrt{\mathsf{x}}$ 
			%\end{equation}
			is strictly decreasing.
			%there exists a $\underline{d} > 0$ such that, for all $d \leq \underline{d}$, the function 
			%\begin{equation}
			%	\widetilde{\beta}_j(d) \coloneqq \frac{k_2\beta_j''}{k_1} \frac{(d + \widetilde{r})\sqrt{d + \widetilde{r}} }{2r_{\mathcal{W}}}
			%\end{equation}
			%is decreasing.
			%$\widetilde{r} \coloneqq \bar{r}_{\textup{d}} + \min_{j\in\mathcal{J}}r_{o_j}$.
		\end{enumerate}	
	\end{definition}
	
	%By using the first property we will guarantee that, by keeping $\beta(d_j(x))$ bounded, there are no collisions with the obstacles or the free space boundary. Property $2$ will be used for the asymptotic stability of the desired point $x=x_{\textup{d}}$. Property $3$ places the rest of the critical points of $\phi$  (which are proven to be saddle points) close to the obstacles, and the last property is used to guarantee that these are non-degenerate.	
	{An example for $\beta$ that satisfies properties 1) and 4), is
%	\begin{equation*}
%		\beta(z) \coloneqq 
%		\begin{cases} \displaystyle
%		\bar{\beta}\frac{\exp\left(-\frac{1}{z}\right) + \exp\left(-\frac{1}{\tau - z}\right)}{\exp\left(-\frac{1}{z}\right)}, &  z \leq \tau \\
%		\bar{\beta}, & z \geq \tau,
%		\end{cases}	 %\label{eq:beta_j_exp}, 
%	\end{equation*}
%	for sufficiently small $\tau$, or the functions
	\begin{align}
	\beta(\mathsf{x}) \coloneqq& 
	\begin{cases} \displaystyle
	(6\mathsf{x}^5 - 15\mathsf{x}^4 + 10\mathsf{x}^3)^{-1}, &  \mathsf{x} \leq 1\\
	1, & \mathsf{x} \geq 1 ,
	\end{cases} \label{eq:beta_exps} 	 %\label{eq:beta_j_exp_pol},  
	%\beta(z) \coloneqq& 
	%\begin{cases} \displaystyle
	%\ln^4\left(\frac{z}{\tau}\right), &  z \leq \tau\\
	%0, & z \geq \tau.
	%\end{cases} \notag
	\end{align}}
	%Note that $\beta'(z) = \beta''(z) = 0$, for $z \geq 1$. 	
%	\vspace{-.5mm}	
	%Note that $\beta$ is essentially a reciprocal barrier function 		\cite{wang2017safety}.
	By appropriately choosing $\tau$, only one $\beta(d_\ell(\chi))$, $\ell\in\bar{\mathcal{J}}_0$ affects the robotic agent for each $\chi\in\mathcal{F}_\mathcal{H}$, with {${\beta'(d_\ell(\chi_\textup{d}))} = {\beta''(d_\ell(\chi_\textup{d}))} = 0$.  Hence, properties 2) and 3) of Def. \ref{def:2nd nf} are satisfied}.
	\begin{proposition}[\cite{verginis2021adaptive}] \label{prop:tau}
	By choosing $\tau$ as $\tau \in (0,\min\{\bar{r}^2, \bar{r}_{\textup{d}}\})$,
	%\begin{equation} \label{eq:tau choice}
	%\tau \in (0,\min\{\bar{r}^2, \bar{r}_{\textup{d}}\}),
	%\end{equation} 		
	we guarantee that at each $\chi \in \mathcal{F}_\mathcal{H}$	 
	{there is at most one $\ell\in\bar{\mathcal{J}}_0$} such that $d_\ell(\chi) \leq \tau$, implying that $\beta'(d_\ell(\chi))$ and $\beta''(d_\ell(\chi))$ are non-zero. 
	\end{proposition}	
    Intuitively, the obstacles and the workspace boundary  have a local region of influence defined by the constant $\tau$.
    
    To compensate for the unknown mass and friction terms of the dynamics \eqref{eq:agent dynamics navigation}, we define the estimates $\hat{m}$, $\hat{\alpha}$ of $m$ and $\alpha$ (see Assumption \ref{ass:friction}), respectively, to be used in the control design.
    
    Given the aforementioned definition, we design a reference signal $v_{\textup{d}}:\mathcal{F}_\mathcal{H}\to\mathbb{R}^2$ for the robot velocity $\dot{x}$ as
	\begin{equation} \label{eq:v_d}
		v_{\textup{d}}(\chi) = -J_\mathcal{H}(x)^{-1} \nabla_{\mathcal{H}(x)} \varphi(\mathcal{H}(x)),
	\end{equation}
	where $J_\mathcal{H}(x) \coloneqq \frac{\partial \mathcal{H}(x)}{\partial x}$ is the nonsingular Jacobian matrix of $\mathcal{H}$. Next, 
	we define the respective velocity error $e_v \coloneqq \dot{x} - 	v_{\textup{d}}(x)$, and design the
	control law as $u:\mathcal{F} \times \mathbb{R}^{4} \to \mathbb{R}^2$, with 
	\begin{multline}
	    u\coloneqq u(x,v,\hat{m},\hat{\alpha}) \coloneqq -k_\varphi J_\mathcal{H}(x)^\top \nabla_{\mathcal{H}(x)} \varphi(\mathcal{H}(x)) + \hat{m}\dot{v}_\text{d} \\ -  \left(k_v + \frac{3}{2}\hat{\alpha}\right)e_v,
	\end{multline}
	where $k_\varphi$ is a positive gain, and $\hat{m}$, $\hat{\alpha}$ evolve according to 
	\begin{subequations} \label{eq:adaptation law}
	\begin{align} 
	    \dot{\hat{m}} \coloneqq& -k_m e_v^\top \dot{v}_{\textup{d}}  \\
		\dot{\hat{\alpha}} \coloneqq & k_\alpha \|e_v\|^2, 
	\end{align}
	\end{subequations}
	with $k_m$, $k_\alpha$ positive constants. 
	The aforementioned control protocol guarantees safe asymptotic convergence of the robot to its goal from almost all collision-free initial conditions, i.e., except for a set of measure zero, provided that $\tau$ is sufficiently small and $k_\varphi > \frac{\alpha}{2}$ (see Theorem 2 in \cite{verginis2021adaptive}). Therefore, since the convergence is asymptotic, we conclude that there exists a finite time instant $t_i > t_0$ such that
	$\bar{\mathcal{B}}(x_i(t_i),r_i) \subset \pi_{i'_k}$, achieving thus the transition $\pi_{i_k} \to_i \pi_{i'_k}$. \\
	
\noindent \textbf{{2) Cooperative Object Transportation}}

We deal now with the control design for the cooperative object transportation problem. 
Consider an object $j \in\mathcal{M}$ grasped by a team of robots $\mathcal{A}$ at $t_0 \in \mathbb{R}_{\geq 0}$, i.e., $\mathcal{AG}_j(t_0) = \mathcal{A}$, evolving subject to the dynamics \eqref{eq:coupled dynamics} and satisfying $\bar{\mathcal{B}}(x^o_j(t_0),r_{\mathcal{A},j}) \subset \pi_{j_k}$, for some $j_k \in \mathcal{K}$. The goal is the transportation of the object to some $\pi_{j_k'}$, $j_k' \in \mathcal{K}$. 
As in the robot navigation case, we let $x^e_j = x_j$, $r^e_j = r^o_{\mathcal{A},j}$, denoting the entity consisting of object $j$ and the robots $\mathcal{A}$, viewing the rest of the entities as static obstacles. By also aiming to avoid entering the RoI $\pi_k$, for $k\in\mathcal{K} \backslash\{j_k,j_k'\}$, the free space of the entity $\mathcal{F}_j = (\mathcal{W}_\textup{f} \backslash \mathcal{O}_j) \ominus r^o_{\mathcal{A},j}$, where 
\begin{align*}
\mathcal{O}_i \coloneqq  \left(\bigcup_{\ell \in\mathcal{T}(\mathcal{AG})\backslash\{j\}} \bar{\mathcal{B}}(x^e_\ell,r^e_\ell) \right)  \bigcup \left( \bigcup_{k\in\mathcal{K}\backslash\{j_k,j_k'\}} \pi_k  \right)
\end{align*}

Let the desired navigation goal of the object be $x_{\textup{d},j}$, provided by the procedure $\mathcal{P}_s$. Similarly to the robot navigation case, we use the transformation $\chi = \mathcal{H}(x)$ to transform the environment to the open unit disk $\mathcal{D}$, the obstacles to points $b_\ell$, the object-robots entity to the point $\chi$, and the object goal to $\chi_\textup{d} = \mathcal{H}(x_{\textup{d},j})$. Next, by employing the function $\beta()$, we design a reference signal $v_\textup{d}:\mathcal{F}_\mathcal{H} \to \mathbb{R}^2$ for the object velocity that is identical to \eqref{eq:v_d}. In addition, we define
the adaptation variables $\hat{m}_{\mathcal{A},j}$ and $\hat{\alpha}_{\mathcal{A},j}$ as the estimates of the unknown coupled mass and friction coefficients $m_{\mathcal{A},j}$ and ${\alpha}_{\mathcal{A},j}$ (see Assumption \ref{ass:friction}), respectively,  and design the control law for the robots $\mathcal{A}$ as $u:\mathcal{F}\times \mathbb{R}^4 \to \mathbb{R}^2$, with 
\begin{align*}
	    u_\ell \coloneqq  & u_\ell(x,v,\hat{m}_{\mathcal{A},j},\hat{\alpha}_{\mathcal{A},j}) \\
	    \coloneqq & - \mathsf{cf}_{\ell} \bigg\{ k_\varphi J_\mathcal{H}(x)^\top \nabla_{\mathcal{H}(x)} \varphi(\mathcal{H}(x)) + \hat{m}_{\mathcal{A},j}\dot{v}_\text{d} \\&
	    -  \left(k_v + \frac{3}{2}\hat{\alpha}_{\mathcal{A},j}\right)e_v \bigg\},
	\end{align*}
	where $\mathsf{cf}_\ell$ are \textit{load sharing coefficients} satisfying $\mathsf{cf}_\ell \geq 0$, for all $\ell\in\mathcal{A}$, and $\sum_{\ell \in \mathcal{A}}\mathsf{cf}_\ell = 1$; $k_\varphi$ is a positive gain, and $\hat{m}_{\mathcal{A},j}$, $\hat{\alpha}_{\mathcal{A},j}$ evolve according to \eqref{eq:adaptation law}.
	
	By invoking the property $\sum_{\ell \in \mathcal{A}}\mathsf{cf}_\ell = 1$ and the fact that the object-robots system is converted to a point by the transformation $\mathcal{H}$, we guarantee, similar to the robot navigation case, the safe, asymptotic convergence of the object-robots entity to $\pi_{j'_k}$ from almost all collision-free initial conditions, provided that $k_\varphi > \frac{\alpha}{2}$. Hence, there exists a finite time instant $t_j > t_0$ such that
	$\bar{\mathcal{B}}(x_j(t_j),r^o_{\mathcal{T},j}) \subset \pi_{j'_k}$, achieving thus the transition $\pi_{i_k} \xrightarrow{T}_{\mathcal{A},j} \pi_{j'_k}$.

\subsection{High-Level Plan Generation} \label{sec:high level plan}

The second part of the solution is the derivation of a high-level plan that satisfies the given LTL formulas $\phi_i$ and $\phi^{o}_j$. Thanks to (i) the proposed control laws that allow robot transitions and object transportations $\pi_k\rightarrow_i\pi_{k'}$ and $\pi_k\xrightarrow{T}_{\mathcal{A},j}\pi_{k'}$, respectively, and (ii) the off-the-self control laws that guarantee grasp and release actions $i\xrightarrow{g}j$ and $i\xrightarrow{r}j$, we can abstract the behavior of the multi-robot-object system using a finite transition system as presented in the sequel.

\begin{definition} \label{def:TS objects all agents}
The coupled behavior of the overall system of all the $N$ agents and $M$ objects is modeled by the transition system $$\mathcal{TS} = (\Pi_s,\Pi^{\text{init}}_s, \rightarrow_{s},\mathcal{AG}, \Psi, \mathcal{L}, \Lambda, {P}_s,\chi)$$
where 
\begin{enumerate}[label=(\roman*), align=left, leftmargin=0pt, listparindent=\parindent, labelwidth=0pt, itemindent=!]
\item $\Pi_s\subset \bar{\Pi}\times\bar{\Pi}^{o}\times (2^{\mathcal{N}_0})^M$ is the set of states; 
$\bar{\Pi}\coloneqq\Pi_1\times\cdots\times\Pi_N$ and $\bar{\Pi}^{o}\coloneqq\Pi^{o}_1\times\cdots\times\Pi^{o}_M$ are the set of states-regions that the agents and the objects can be at, with $\Pi_i  = \Pi^{o}_j = \Pi,\forall i\in\mathcal{N},j\in\mathcal{M}$; 
%$\mathcal{AG}$ is the set of grasping variables introduced in Section \ref{sec:Model and PF}.
%, with
%$\mathcal{AG}_i \coloneqq \{\mathcal{AG}_{i,0}\}\cup\{[\mathcal{AG}_{i,j}]_{j\in\mathcal{M}}\}, \forall i\in\mathcal{N}$. 

By defining 
$\bar{\pi} \coloneqq \left(\pi_{k_1},\cdots,\pi_{k_N}\right),\bar{\pi}_{o}  \coloneqq (\pi_{\scriptscriptstyle k^{o}_1},\cdots,\pi_{\scriptscriptstyle k^{o}_M})$, %\bar{w}=\left(w_1,\cdots,w_M\right)$, with $\pi_{k_i},\pi_{k^{o}_j}\in\Pi$ (i.e., $k_i,k^{o}_j\in\mathcal{K},\forall i\in\mathcal{N},j\in\mathcal{M}$) 
%and $w_i\in\mathcal{AG}_i, \forall i\in\mathcal{N}$, 
then the coupled state $\pi_s \coloneqq (\bar{\pi},\bar{\pi}_{o},\mathcal{AG})$ belongs to $\Pi_s$, i.e., $(\bar{\pi},\bar{\pi}_{o},\mathcal{AG})\in\Pi_s$ if
\begin{enumerate}
\item $\mathcal{P}_{s,0}\Big(r_{\pi_k}, [r_i]_{i\in\{ i\in\mathcal{N}: k_i = k \} }, [r^{o}_j]_{j\in\{ j\in\mathcal{M}:k^{o}_j= k \} }\Big) = \top$, i.e., the respective robots and objects fit in the region, $\forall k\in\mathcal{K}$, 
\item $k_i = k^{o}_j$ for all $i\in\mathcal{N}, j\in\mathcal{M}$ such that $i \in \mathcal{AG}_{j}$, i.e., a robot must be in the same region with the object it grasps,
\end{enumerate} 
\item $\Pi^{\text{init}}_s\subset\Pi_s$ is the initial set of states at $t=0$, which, owing to (i), satisfies the conditions of Problem \ref{problem},\\
\item $\rightarrow_s\subset \Pi_s\times\Pi_s$ is a transition relation defined as follows: given the states $\pi_s, \widetilde{\pi}_s\in\Pi$, with
\begin{align}
\pi_s \coloneqq & (\bar{\pi},\bar{\pi}_{o}, \mathcal{AG}) \notag \\
\coloneqq & (\pi_{k_1},\dots,\pi_{k_N}, \pi_{k^{o}_1},\dots,\pi_{k^{o}_M},  \mathcal{AG}_1,\dots, \mathcal{AG}_M), \notag \\
\widetilde{\pi}_s \coloneqq & ( \widetilde{\bar{\pi}},{\widetilde{\bar{\pi}}}_{o}, \widetilde{\mathcal{AG}} ) \notag \\ \coloneqq & (\pi_{\widetilde{k}_1},\dots,\pi_{\widetilde{k}_N}, \pi_{\widetilde{k}^{o}_1},\dots,\pi_{\widetilde{k}^{o}_1},\widetilde{\mathcal{AG}}_1,\dots, \widetilde{\mathcal{AG}}_M), \label{eq:pi_s}
\end{align}
a transition $\pi_s \rightarrow_s {\pi}'_s $ occurs if all the following hold:

\begin{enumerate}
	\item $\nexists i\in\mathcal{N}, j\in\mathcal{M}$ such that $i \in \mathcal{AG}_j$, $i\notin \widetilde{\mathcal{AG}}_j$ (or $i \notin \mathcal{AG}_j$, $i\in \widetilde{\mathcal{AG}}_j$), 
	and $k_i \neq \widetilde{k}_i$, i.e., there are no simultaneous grasp/release and navigation actions,
	\item $\nexists i\in\mathcal{N}, j\in\mathcal{M}$ such that $i \in \mathcal{AG}_j$, $i\notin \widetilde{\mathcal{AG}}_j$ (or $i \notin \mathcal{AG}_j$, $i\in \widetilde{\mathcal{AG}}_j$), and $k_i = k^{o}_j \neq \widetilde{k}_i=\widetilde{k}^{o}_j$, i.e., there are no simultaneous grasp/release and transportation actions,
	\item $\nexists i\in\mathcal{N}, j,j'\in\mathcal{M}$, with $j\neq j'$, such that $i \in \mathcal{AG}_j$ and $i \in \widetilde{\mathcal{AG}}_{j'}$,	i.e., there are no simultaneous grasp and release actions,
	\item $\nexists j\in\mathcal{M}$ such that $k^{o}_j \neq \widetilde{k}^{o}_j$ and $i \notin \mathcal{AG}_j, \forall i\in\mathcal{N}$ ( or $i \notin \widetilde{\mathcal{AG}}_{j}, \forall i\in\mathcal{N}$), i.e., there is no transportation of a non-grasped object,
	\item $\nexists j\in\mathcal{M},\mathcal{T}\subseteq \mathcal{N}$ such that $k^{o}_j \neq \widetilde{k}^{o}_j$ and $\Lambda(m^{o}_j, \zeta_{\mathcal{T}}) = \bot$, with $(i \in \mathcal{AG}_j, i \in \widetilde{\mathcal{AG}}_j ) \Leftrightarrow i\in\mathcal{T}$, i.e., the agents grasping an object are powerful enough to transfer it,
\end{enumerate}

\item $\Psi \coloneqq \bar{\Psi}\cup\bar{\Psi}^{o}$ with $\bar{\Psi}=\bigcup_{i\in\mathcal{N}}\Psi_{i}$ and $\bar{\Psi}^{o} = \bigcup_{j\in\mathcal{M}}\Psi^{o}_j$, are the atomic propositions of the agents and objects, respectively, as defined in Section \ref{sec:Model and PF}.

\item $\mathcal{L}:\Pi_s \rightarrow 2^\Psi$ is a labeling function defined as follows: Given a state $\pi_s$ as in \eqref{eq:pi_s} and $\psi_s \coloneqq \Big( \bigcup_{i\in\mathcal{N}}\psi_i\Big)\bigcup \Big(\bigcup_{j\in\mathcal{M}}\psi^{o}_j\Big)$ with $\psi_i\in2^{\Psi_i},\psi^{o}_j\in2^{\Psi^{o}_j}$, then $\psi_s\in\mathcal{L}(\pi_s)$ if  $\psi_i\in\mathcal{L}_i(\pi_{k_i})$ and $\psi^{o}_j\in\mathcal{L}^{o}_j(\pi_{k^{o}_j}), \forall i\in\mathcal{N},j\in\mathcal{M}$.

\item $\Lambda$ and ${P}_s$ as defined in Section \ref{sec:Model and PF}.

\item $\chi: (\to_s) \to \mathbb{R}_{\geq 0}$ is a function that assigns a cost to each transition $\pi_s \to_s \widetilde{\pi}_s$. This cost might be related to the distance of the robots' regions in $\pi_s$ to the ones in $\widetilde{\pi}_s$, combined with the cost efficiency of the robots involved in transport tasks (according to $\zeta_i, i\in\mathcal{N}$).
\end{enumerate} 
\end{definition}

%Broadly speaking, the cases (1)-(4) in the transition relation of Def. \ref{def:TS objects all agents} correspond to agent transition, object transportation, grasp and release actions, respectively.  More specifically, the agents $z\in\mathcal{Z}$ perform transitions between $\pi_{k_z}, \pi_{k'_z}$, the agents $v\in\mathcal{V}$ perform transportation of the objects $s_v\in\mathcal{S}$ from $\pi_{k_v}$ to $\pi_{k'_v}$, and the agents $g\in\mathcal{G}, q\in\mathcal{Q}$ perform grasping and releasing actions with objects $x_g\in\mathcal{X}$ and $y_q\in\mathcal{Y}$, respectively. 

Next, we form the global LTL formula $\phi \coloneqq (\land_{i\in\mathcal{N}}\phi_i)\land(\land_{j\in\mathcal{M}}\phi^{o}_j)$ over the set $\Psi$.
Given$\phi$, we define the \textit{language} $\texttt{Words}(\phi)=\left\{\sigma\in (2^{\Psi})^{\omega}|\sigma\models\phi\right\}$, where $\models\subseteq (2^\Psi)^{\omega}\times\phi$ is the satisfaction relation, as the set of infinite words $\sigma\in (2^{\Psi})^{\omega}$ that satisfy the LTL formula $\phi$. Then, we translate $\phi$ to a B\"uchi Automaton $\mathcal{BA}$ defined over $(2^\Psi)^{\omega}$. The B\"uchi Automaton is defined as follows \cite{baier2008principles}:

\begin{definition}[NBA]\label{def:nba}
A \textit{Nondeterministic B$\ddot{\text{u}}$chi Automaton} (NBA) $\mathcal{BA}$ is defined as a tuple $B=\left(\Pi_{B}, \Pi_{B}^0,\Sigma,\rightarrow_B,\Pi_B^F\right)$, where $\Pi_{B}$ is the set of states, $\Pi_{B}^0\subseteq\Pi_{B}$ is a set of initial states, $\Sigma=2^{\Psi}$ is the alphabet, $\rightarrow_{B}\subseteq\Pi_{B}\times \Psi\times\Pi_{B}$ is the transition relation, and $\Pi_B^F\subseteq\Pi_{B}$ is a set of accepting/final states.
\end{definition}

Given the $\mathcal{TS}$ and the NBA $\mathcal{BA}$ that corresponds to the LTL $\phi$, we can now define the \textit{Product B$\ddot{\text{u}}$chi Automaton} (PBA) $\mathcal{PBA} \coloneqq \mathcal{TS}\otimes \mathcal{BA}$, as follows \cite{baier2008principles}:

\begin{definition}[PBA]\label{defn:pba}
Given the product transition system $\mathcal{TS} = (\Pi_s,\Pi^{\text{init}}_s, \rightarrow_{s},\mathcal{AG}, \Psi, \mathcal{L}, \Lambda, {P}_s,\chi)$ and the NBA $B=\left(\Pi_{B}, \Pi_{B}^0,\Psi,\rightarrow_B,\Pi_B^F\right)$, we can define the \textit{Product B$\ddot{\text{u}}$chi Automaton} $\mathcal{PBA} \coloneqq \mathcal{PTS}\otimes \mathcal{BA}$ as a tuple $\mathcal{PBA}=(\Pi_P, \Pi_P^0,\longrightarrow_{P},\Pi_P^F)$ where (a) $\Pi_P=\Pi_{\text{s}}\times\Pi_{B}$ is the set of states; (b) $\Pi_P^0=\Pi_s^{\text{init}}\times\Pi_B^0$ is a set of initial states; (c) $\longrightarrow_{P}\subseteq\Pi_P\times \Sigma \times\Pi_P$ is the transition relation defined by the rule: $\frac{(\pi_s\rightarrow_s \pi_s')\wedge( \pi_{B}\xrightarrow{\mathcal{L}\left(\pi_s\right)}\pi_{B}')}{\pi_{P}=\left(\pi_s,\pi_{B}\right)\longrightarrow_P \pi_{P}'=\left(\pi_s',\pi_{B}'\right)}$. Transition from state $\pi_P\in\Pi_P$ to $\pi_P'\in\Pi_P$, is denoted by $(\pi_P,\pi_P')\in\longrightarrow_P$, or $\pi_P\longrightarrow_P \pi_P'$; (d) $\chi_P(\pi_{\text{P}},\pi_{\text{P}}')=\chi(\pi_s,\pi_s')$,  where $\pi_{P}=(\pi_s,\pi_B)$ and $\pi_{\text{P}}'=(\pi_{s}',\pi_B')$; and (e) $\Pi_P^F=\Pi_{s}\times\Pi_B^F$ is a set of accepting/final states. 
\end{definition} 

Given $\phi$ and the PBA, an infinite path $\pi_{\text{pl}}\coloneqq \pi_{s,1} \pi_{s,2}\dots$ of a $\mathcal{TS}$ satisfies $\phi$ if and only if $\texttt{trace}(\pi_{\text{pl}})\in\texttt{Words}(\phi)$, which is equivalently denoted by $\pi_{\text{pl}}\models\phi$, where $\texttt{trace}(\pi_{\text{pl}})\in\left(2^{\Psi}\right)^{\omega}$ is defined as 
$\texttt{trace}(\pi_{\text{pl}}) = \mathcal{L}(\pi_{s,1})\mathcal{L}(\pi_{s,2})\dots$. Specifically, if there is a path satisfying $\phi$, then there exists a path  $\pi_{\text{pl}}\models\phi$ that can be written in a finite representation, called prefix-suffix structure, i.e., $\pi_{\text{pl}}=\pi_{\text{pl}}^{\text{pre}}[\pi_{\text{pl}}^{\text{suf}}]^{\omega}$, where the prefix part $\pi_{\text{pl}}^{\text{pre}}$ is executed only once followed by the indefinite execution of the suffix part $\pi_{\text{pl}}^{\text{suf}}$. The prefix part $\pi_{\text{pl}}^{\text{pre}}$ is the projection of a finite path $p^{\text{pre}}$ that lives in $\Pi_P$ onto $\Pi_{s}$. 
%The path $p^{\text{pre}}$ starts from an initial state $q_P^0\in\ccalQ_P^0$ and ends at a final state $q_P^F\in\ccalQ_P^F$, i.e., it has the following structure $p^{\text{pre}}=(q_{\text{PTS}}^0,q_B^0)(q_{\text{PTS}}^1,q_B^1)\dots (q_{\text{PTS}}^K,q_B^K)$ with $(q_{\text{PTS}}^K,q_B^K)\in\ccalQ_P^F$. The suffix part $\tau^{\text{suf}}$ is the projection of a finite path $p^{\text{suf}}$ that lives in $\ccalQ_P$ onto $\ccalQ_{\text{PTS}}$. The path $p^{\text{suf}}$ is a cycle around the final state $(q_{\text{PTS}}^K,q_B^K)$, i.e., it has the following structure $p^{\text{suf}}=(q_{\text{PTS}}^{K},q_B^K)(q_{\text{PTS}}^{K+1},q_B^{K+1})\dots (q_{\text{PTS}}^{K+S},q_B^{K+S})(q_{\text{PTS}}^{K+S+1},q_B^{K+S+1})$, where $(q_{\text{PTS}}^{K+S+1},q_B^{K+S+1})=(q_{\text{PTS}}^{K},q_B^{K})$. Then our goal is to compute a plan $\tau=\tau^{\text{pre}}[\tau^{\text{suf}}]^{\omega}=\Pi|_{\text{PTS}}p^{\text{pre}}[\Pi|_{\text{PTS}}p^{\text{pre}}]^{\omega}$, where $\Pi|_{\text{PTS}}$ stands for the projection on the state-space $\ccalQ_{\text{PTS}}$,

Computing a plan $\pi_{\text{pl}}$ is typically accomplished by applying graph-search methods to the PBA. Specifically, to generate a motion plan $\pi_{\text{pl}}$ that satisfies $\phi$, the PBA is viewed as a weighted directed graph $\mathcal{G}_P=\{\mathcal{V}_P, \mathcal{E}_P, w_P\}$, where the set of nodes $\mathcal{V}_P$ is indexed by the set of states $\Pi_P$, the set of edges $\mathcal{E}_P$ is determined by the transition relation $\longrightarrow_P$, and the weights assigned to each edge are determined by the function $\chi_P$. Then, to find the optimal plan $\tau\models\phi$, shortest paths towards final states and shortest cycles around them are computed. More details about this approach can be found in \cite{smith2011optimal,ulusoy2014optimal} and the references therein. {While any of the aforementioned methodologies could be used, in this work we employ $\text{STyLuS}^*$, an algorithm that is designed to solve complex temporal planning problems in large-scale multi-robot systems and has been shown to achieve significantly lower complexity, in terms of running time and memory, than standard graph-search methods 
\cite{kantaros2020stylus}.}
%Nevertheless, these graph-based methods scale poorly as they require to construct the PBA whose state space increases exponentially as the number of robots or objects and the state space of $\mathcal{TS}$ and $\mathcal{BA}$ increase. 
%To enhance the scalability of the proposed planning algorithm, we employ $\text{STyLuS}^*$, an algorithm that is designed to solve complex temporal planning problems in large-scale multi-robot systems \cite{kantaros2020stylus}. 
Specifically, $\text{STyLuS}^*$ is a sampling-based method that builds incrementally trees that approximate the state-space and transitions of the product automaton and does not require sophisticated graph-search techniques. Technically, $\text{STyLuS}^*$ builds a tree $\mathcal{G}_T=\{\mathcal{V}_T,\mathcal{E}_T,\texttt{Cost}\}$ first,  where $\mathcal{V}_T\subseteq\Pi_P$ is the set of nodes, $\mathcal{E}_T$ is the set of edges, and $\texttt{Cost}$ is defined as per $\chi_P$, determining the cost of reaching a tree node from its root. This tree is rooted at the initial state $\pi_P^0$ and is used for the synthesis of the prefix part. The tree is constructed incrementally, in a sampling-based fashion, and its construction terminates after a user-specified number of iterations $n_{\text{max}}$. Then, we compute paths in the constructed tree structure that connect the root to the detected, if any, final states. These paths correspond to prefix parts of candidate feasible paths. To construct the corresponding suffix paths, new trees are built, similarly, rooted at the previously detected final states aiming to compute cycles around the tree roots. Among all the detected prefix-suffix paths, $\text{STyLuS}^*$ returns the one with the minimum cost. As it was shown in \cite{kantaros2020stylus}, $\text{STyLuS}^*$ is probabilistically complete and asymptotically optimal; that is, the probability of finding a feasible and the optimal solution converges to $1$ as $n_{\text{max}}\to\infty$.

Finally, note that the constructed trees explore the state space of the product automaton and, therefore, the designed prefix-suffix path is defined as an infinite sequence of product automaton states. By projecting it onto the state-space of the transition system $\mathcal{TS}$,  we obtain a high-level prefix-suffix plan defined as a sequence of states $\pi_\text{pl} \coloneqq \pi_{s,1} \pi_{s,2}\dots \models \phi$. The corresponding sequence of atomic propositions is $\psi_\text{pl} =\texttt{trace}(\pi_\text{pl})=\psi_{s,1} \psi_{s,2}\dots  $, with  
\begin{align*}
& \pi_{s,\ell} \coloneqq (\bar{\pi}_\ell,\bar{\pi}_{o,\ell},\bar{w}_\ell ) \in\Pi_s, \forall \ell\in\mathbb{N}, \notag \\
& \psi_{s,\ell} \coloneqq \Big(\bigcup\limits_{i\in\mathcal{N}}\psi_{i,\ell} \Big)\bigcup\Big(\bigcup\limits_{j\in\mathcal{M}}\psi^{o}_{j,\ell} \Big) \in 2^{\Psi}, \mathcal{L}(\pi_{s,\ell}), \forall \ell\in\mathbb{N},
\end{align*}
where 
\begin{itemize}
\item $\bar{\pi}_\ell \coloneqq \pi_{k_{1,\ell}}, \pi_{k_{2,\ell}}, \dots$ with $k_{i,\ell} \in\mathcal{K},\forall i\in\mathcal{N}$, 
\item $\bar{\pi}_{o,\ell} \coloneqq \pi_{k^{o}_{1,\ell}}, \pi_{k^{o}_{2,\ell}}, \dots$ with $k^{o}_{j,\ell} \in\mathcal{K},\forall j\in\mathcal{M}$, 
\item $\bar{w}_\ell \coloneqq w_{1,\ell}, w_{2,\ell},\dots$ with $w_{i,\ell} \in\mathcal{AG}_i, \forall i\in\mathcal{N}$,
\item $\psi_{i,\ell} \in 2^{\Psi_i}, \mathcal{L}_i(\pi_{k_{i,\ell}}), \forall i\in\mathcal{N}$, 
\item $\psi^{o}_{j,\ell}\in 2^{\Psi^{o}_j}, \mathcal{L}^{o}_j(\pi_{k^{o}_{j,\ell}}), \forall j\in\mathcal{M}$.
\end{itemize}
 
%Since $\Psi_i\cap\Psi_n = \Psi_i\cap\Psi_{o_j} = \Psi_{o_j}\cap\Psi_{o_\ell} = \emptyset, \forall i,n\in\mathcal{N},j,\ell\in\mathcal{M}$ for $i\neq n, j\neq\ell$, i.e. the atomic propositions are disjoint for the agents and the objects, we can project $\psi^m$ as
%$\psi^m = (\cup_{i\in\mathcal{N}}\psi^m_i)\cup(\cup_{j\in\mathcal{M}}\psi^m_{o_j})$, with $\psi^m_i\in2^{\Psi_i}, \psi^m_{o_j}\in2^{\Psi_{o_j}}, \forall m\in\mathbb{N}$. 
The path $\pi_\text{pl}$ is then projected to the individual sequences of the regions $\pi_{k^{o}_{j,1}} \pi_{k^{o}_{j,2}} \dots$ for each object $j\in\mathcal{M}$, as well as to the individual sequences of the regions $\pi_{k_{i,1}} \pi_{k_{i,2}} \dots$ and the boolean grasping variables $w_{i,1} w_{i,2}\dots$ for each robot $i\in\mathcal{N}$. The aforementioned sequences determine the behavior of robot $i\in\mathcal{N}$, i.e., the sequence of actions (transition, transportation, grasp, release or stay idle) it must take.

By the definition of $\mathcal{L}$ in Def. \ref{def:TS objects all agents}, we obtain that $\psi_{i,\ell} \in\mathcal{L}_i(\pi_{k_{i,\ell}}), \psi^{o}_{j,\ell} \in\mathcal{L}^{o}_j( \pi_{k^{o}_{j,\ell}}), \forall i\in\mathcal{N},j\in\mathcal{M}, \ell\in\mathbb{N}$. Therefore, since $\phi = (\land_{i\in\mathcal{N}}\phi_i)\land(\land_{j\in\mathcal{M}}\phi_{o_j})$ is satisfied by $\psi$, we conclude that $\psi_{i,1}\psi_{i,2}\dots \models \phi_i$ and $\psi^{o}_{j,1}\psi^{o}_{j,2}\dots\models \phi^{o}_j, \forall i\in\mathcal{N}, j\in\mathcal{M}$.

%Therefore, we project the path $(p,\psi)$ to the individual paths $\pi^1_{s_i}\pi^2_{s_i}$ of $\mathcal{TS}_i$, and we obtain sequences of states $\pi^1_{s_i}\pi^2_{s_i}\dots$, with $\pi^m_{s_i}\in\Pi_s$, i.e., $\pi^1_{i}\pi^2_{i}\dots$ and $\pi^1_{o_j}\pi^2_{o_j}\dots$ such that 
%\begin{enumerate}[label=(\roman*)]
%\item $\pi^m_i\in\Pi_i = \Pi$ and $\psi^m_i\in\mathcal{L}(\pi^m_i)$,
%\item $\pi^m_{o_j}\in\Pi_{o_j} = \Pi$ and $\psi^m_{o_j}\in\mathcal{L}(\pi^m_{o_j})$,
%\end{enumerate}
%$\forall m\in\mathbb{N}, i\in\mathcal{N},j\in\mathcal{M}$. Note that since $\phi = (\land_{i\in\mathcal{N}}\phi_i)\land(\land_{j\in\mathcal{M}}\phi_{o_j})$ is satisfied by $\psi^m = (\cup_{i\in\mathcal{N}}\psi^m_i)\cup(\cup_{j\in\mathcal{M}}\psi^m_{o_j})$, then $\phi_i$ and $\phi_{o_j}$ are satisfied by $\psi_i = \psi^1_i\psi^2_i\dots$ and $\psi_{o_j} = \psi^1_{o_j}\psi^2_{o_j}\dots$, respectively, $\forall \in\mathcal{N},j\in\mathcal{M}$.

The sequences $\pi_{k_{i,1}}\pi_{k_{i,2}}\dots$, $\psi_{i,1}\psi_{i,2}\dots$ and $\pi_{k^{o}_{j,1}}\pi_{k^{o}_{j,2}}\dots, \psi^{o}_{j,1}\psi^{o}_{j,2}\dots$ over $\Pi, 2^{\Psi_i}$ and $\Pi, 2^{\Psi^{o}_j}$, respectively, produce the trajectories ${q}_i(t)$ and ${x}^{o}_j(t), \forall i\in\mathcal{N},j\in\mathcal{M}$. The corresponding behaviors are $\beta_i = ({q}_i(t),\sigma_i) = ({q}_i(t_{i,1}),\sigma_{i,1})({q}_i(t_{i,2}),\sigma_{i,2})\dots$ and $\beta^{o}_j$ $=$ $({x}^{o}_j(t),\sigma^{o}_j)= ({x}^{o}_j(t^{o}_{j,1}),\sigma^{o}_{j,1})({x}^{o}_j(t^{o}_{j,2}),\sigma^{o}_{j,2})\dots$, respectively, according to Section \ref{subsec:pf}, with $\mathcal{A}_i({q}_i(t_{i,\ell}))\subset \pi_{k_{i,\ell}}, \sigma_{i,\ell}\in\mathcal{L}_i(\pi_{k_{i,\ell}})$ and $\mathcal{O}_j({x}_{o_j}(t_{o_{j,m}}))\in\pi_{k^{o}_{j,\ell}}, \sigma^{o}_{j,\ell} \in\mathcal{L}^{o}_j(\pi_{k^{o}_{j,\ell}})$. Thus, it is guaranteed that $\sigma_i \models \phi_i,\sigma^{o}_j \models \phi^{o}_j$ and consequently, the behaviors $\beta_i$ and $\beta^{o}_j$ satisfy the formulas $\phi_i$ and $\phi^{o}_j$, respectively, $\forall i\in\mathcal{N},j\in\mathcal{M}$. The aforementioned reasoning is summarized in the next theorem:
\begin{theorem}
The execution of the path $(\pi_{\text{pl}},\psi_{\text{pl}})$ of $\mathcal{TS}$ guarantees behaviors $\beta_i,\beta^{o}_j$ that yield the satisfaction of $\phi_i$ and $\phi^{o}_j$, respectively, $\forall i\in\mathcal{N},j\in\mathcal{M}$, providing, therefore, a solution to Problem \ref{problem}.  
\end{theorem}

%\begin{remark}
%Note that although the overall set of states of $\mathcal{TS}$ increases exponentially with respect to the number of agents/objects/regions, some states are not reachable, due to our constraints for the object transportation and the size of the regions, reducing thus the state complexity.
%\end{remark}

\section{Simulations} \label{sec:simulations}

In this section, we provide two case studies in an obstacle-cluttered office environment. We choose the atomic propositions for the robots and objects as $\Psi_i = \{``i\text{-}\pi_1",\dots,``i\text{-}\pi_K"\}$ and $\Psi^{o}_j = \{``O_j\text{-}\pi_1",\dots,``O_j\text{-}\pi_K"\}$, respectively, for $i\in\mathcal{N}$, $j\in\mathcal{M}$, indicating their presence in the regions of interest. For the constructed transition systems, we set the cost $\chi$ as the Euclidean distance among the RoI contained in the nodes of the transitions.  

For the continuous control design, we choose the robot dynamics of the form \eqref{eq:agent dynamics navigation} with mass $m_i =1$, $f_i(\cdot) = \frac{5}{4} \sin(0.5(x_{i_1} + x_{i_2})) F(\dot{x}_i)\dot{x}_i$, with $F(\dot{x}_i) = \textup{diag}\{ [\exp(-\text{sgn}(\dot{x}_i)\dot{x}_i)+1]_{i\in\{1,2\}}  \}$, where we denote $(x_{i_1},x_{i_2}) = x_i$, $(\dot{x}_{i_1},\dot{x}_{i_2}) = \dot{x}_i$, $i\in\mathcal{N}$. We further choose $\bar{r} = 0.1$ and a variation of \eqref{eq:beta_exps} for $\beta$ with $\tau = \bar{r}^2$. Finally, we set the control gains to $k_1 = 0.01$, $k_2 = 5$, $k_\phi=1$, and $k_m = k_\alpha = 0.01$.

%that illustrate the proposed algorithm. 

\subsection{Case Study I:} 
\begin{figure}
    \centering
    \includegraphics[width=.475\textwidth]{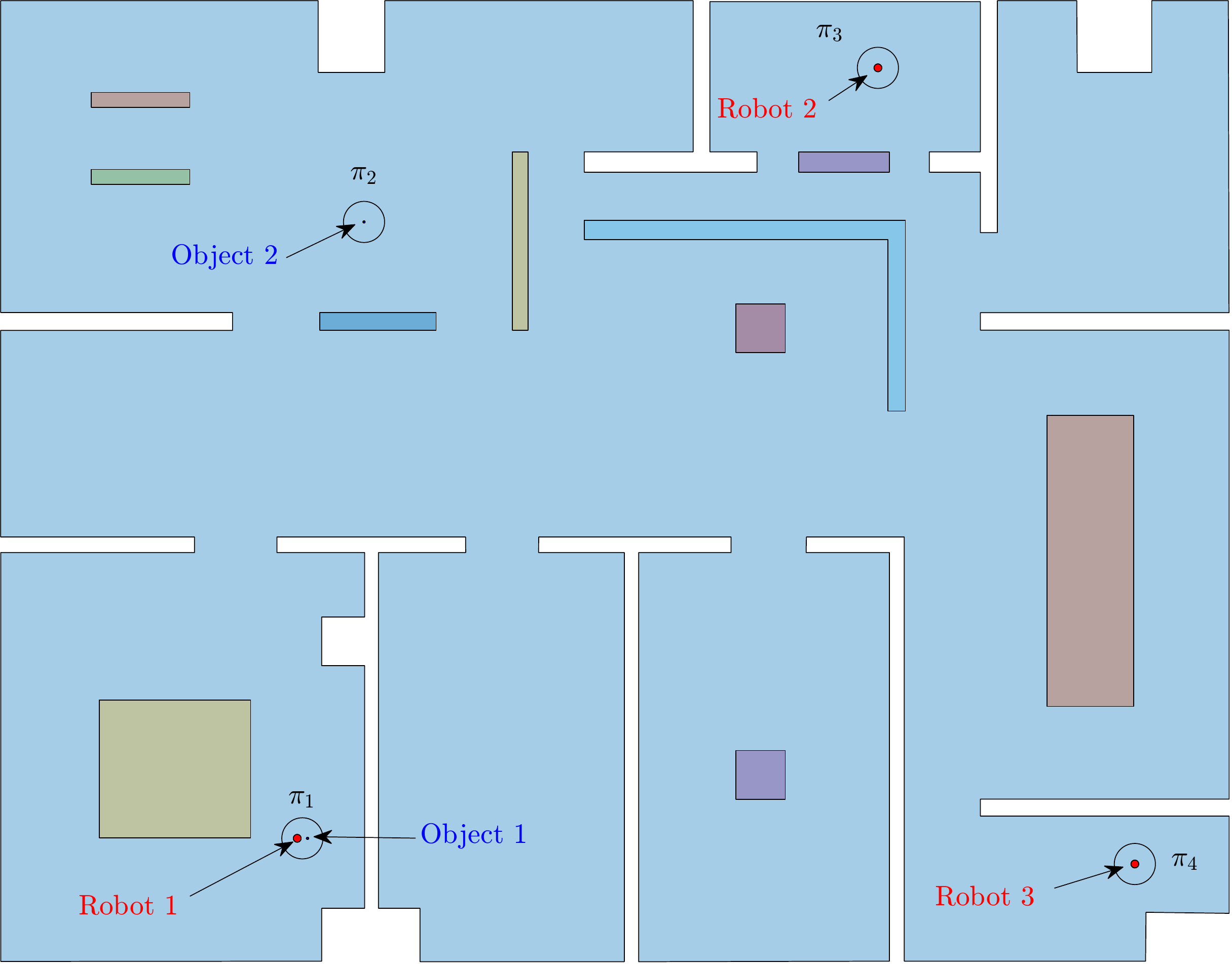}
    \caption{The multi-robot-object initial configuration in the first case study.}
    \label{fig:sim1_init}
\end{figure}

In this case study, we consider $N=3$ robots, $K=4$ regions and $M=2$ objects. The regions of interest are circular centered at $(88,-280)$m, $(100,-160)$m, $(200, -130)$m, and $(250, -285)$m, with radius equal to $4$m.
The robots' and object' mass is taken as $1$kg and $0.25$kg, respectively, while their spherical volume's radii as $0.75$m and $0.2$m, respectively. The power capabilities of the robots are $2, 3, 4$, respectively, and the power required for each object is $5, 6$, respectively. 
Initially, the robots are located in regions $\pi_{init(1)} = \pi_1$, $\pi_{init(2)}=\pi_3$, and $\pi_{init(3)}=\pi_4$, respectively, whereas the objects are located in regions $\pi_{init_o(1)}=\pi_2$, $\pi_{init_o(2)}=\pi_1$, respectively. 
    Fig. \ref{fig:sim1_init} depicts the considered environment and the initial configuration of the multi-robot-object system. 

\begin{figure}
    \centering
    \includegraphics[width=.2375\textwidth]{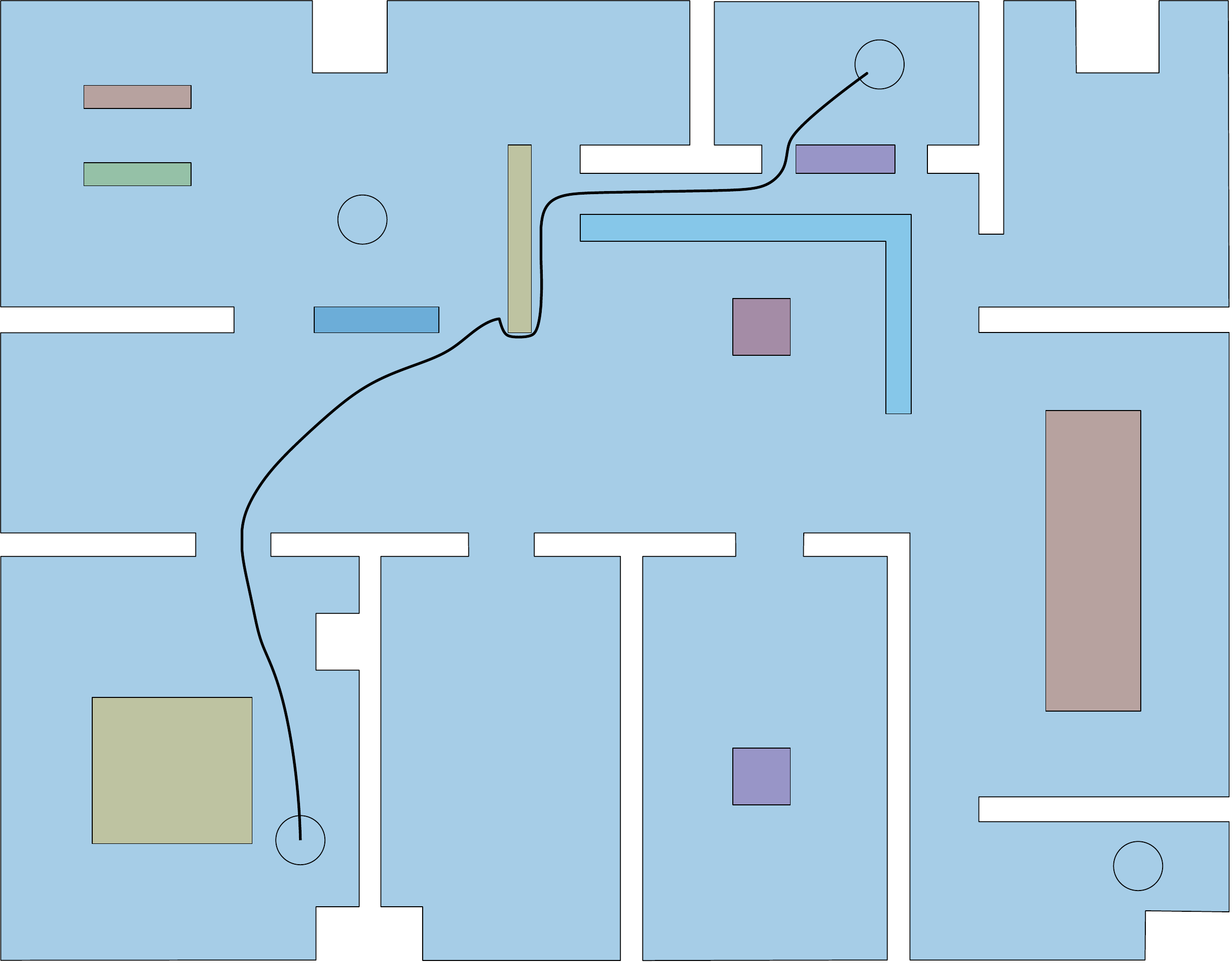}
    \includegraphics[width=.2425\textwidth]{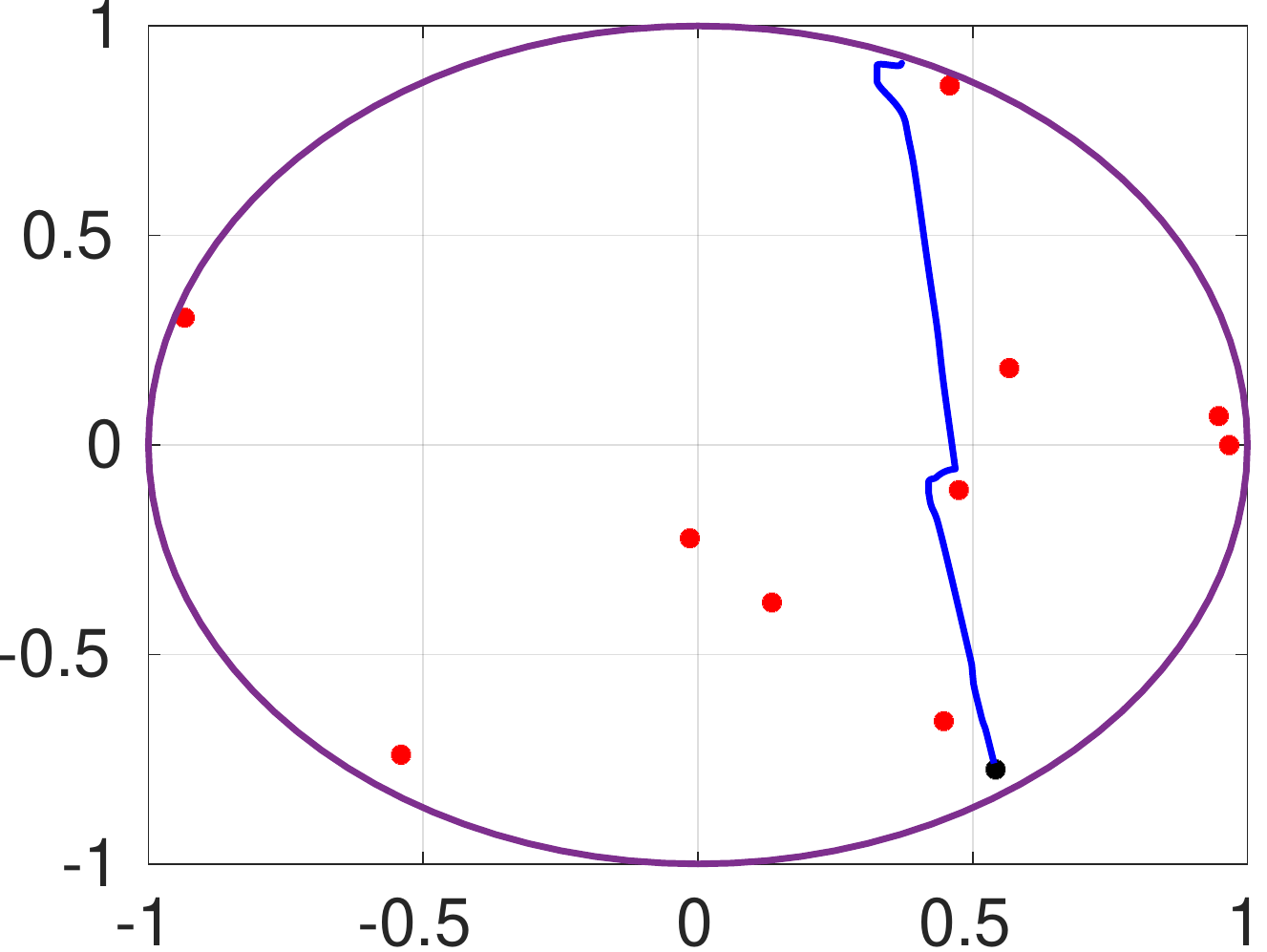}
    \caption{Navigation of robot 1 from $\pi_1$ to $\pi_3$ in the original (left) and the transformed point-world (right) environment. In the original environment, the robot's circular volume has been transferred to the obstacles and workspace boundary.}
    \label{fig:navigation_ex}
\end{figure}

\begin{figure}
    \centering
    \includegraphics[width=.24\textwidth]{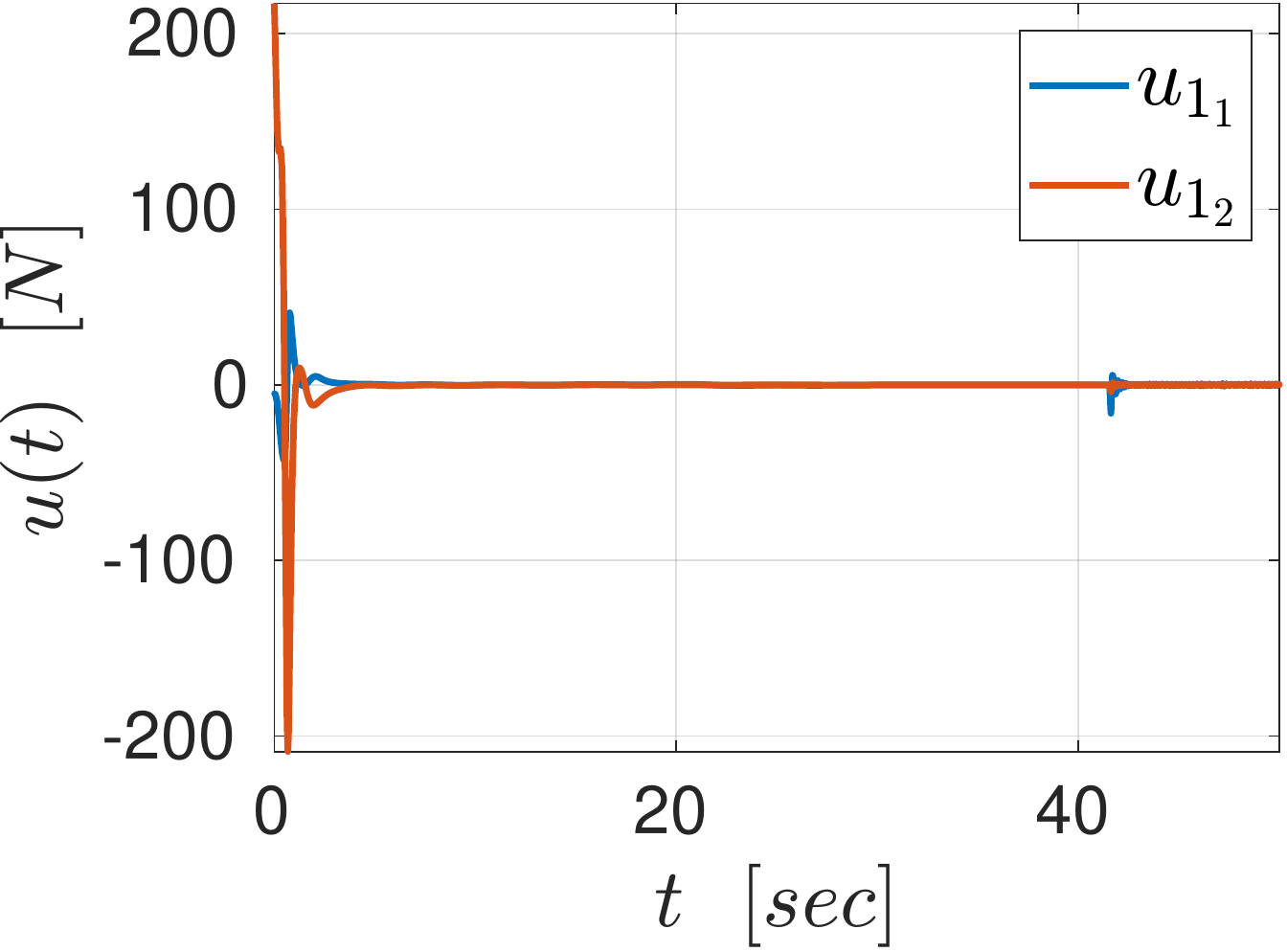}
    \includegraphics[width=.235\textwidth]{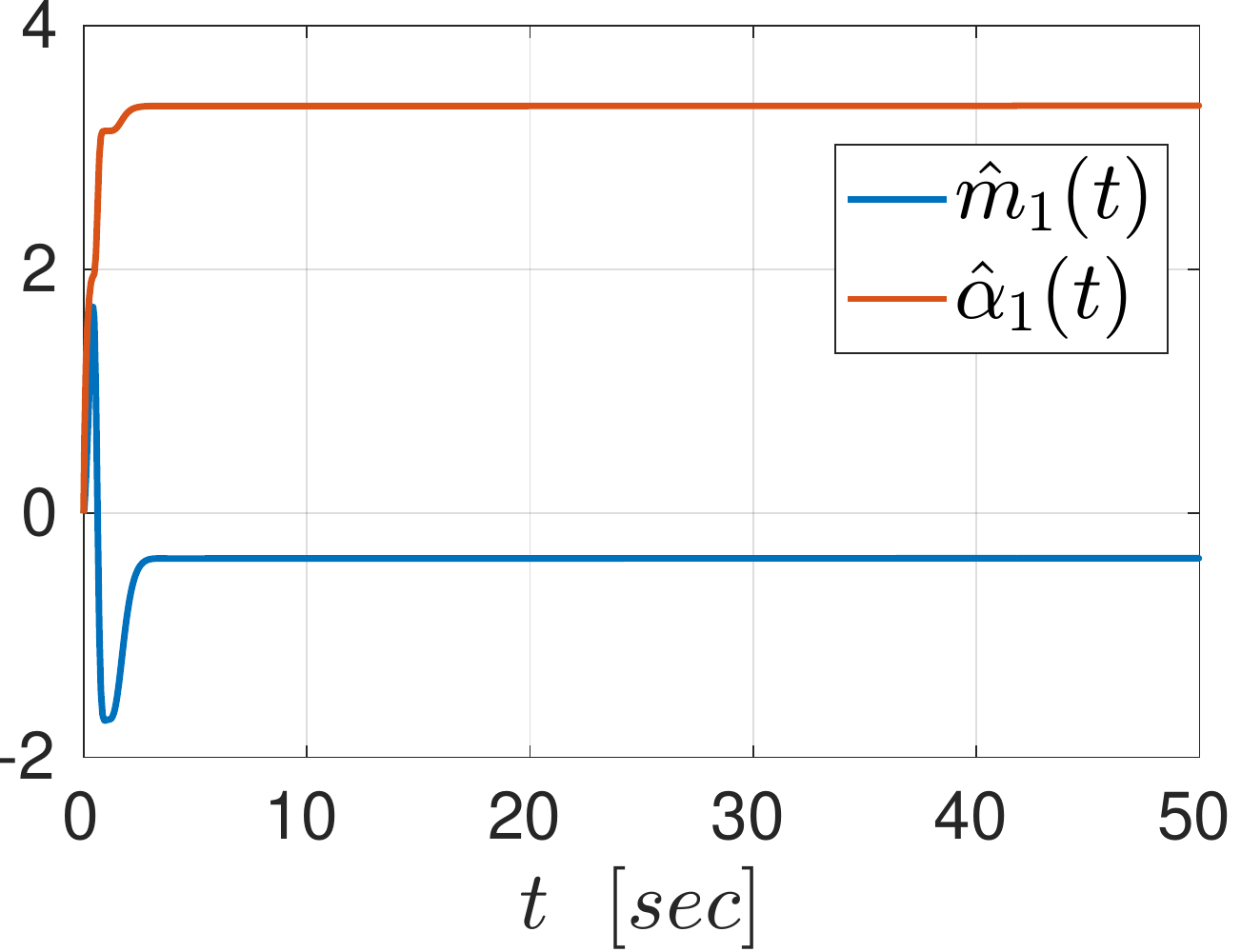}
    \caption{The evolution of the control inputs $u_1(t)$ (left) and adaptation signals $\hat{m}_1(t)$, $\hat{\alpha}_1(t)$ (right) of robot's 1 navigation. }
    \label{fig:navigation_signals}
\end{figure}

\begin{figure}
    \centering
    \includegraphics[width=.2375\textwidth]{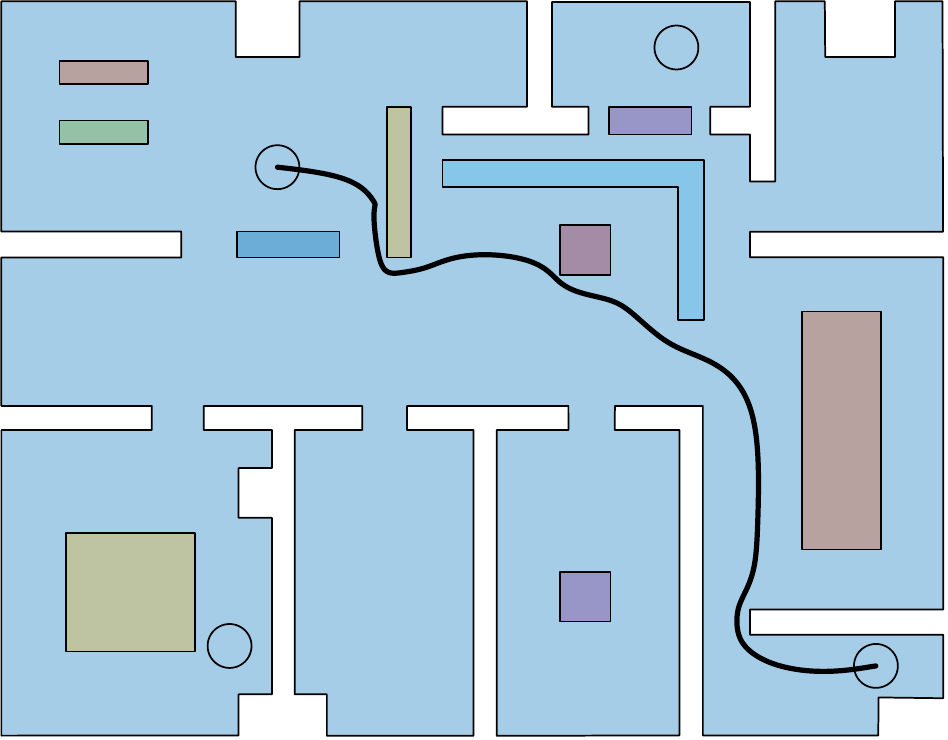}
    \includegraphics[width=.2425\textwidth]{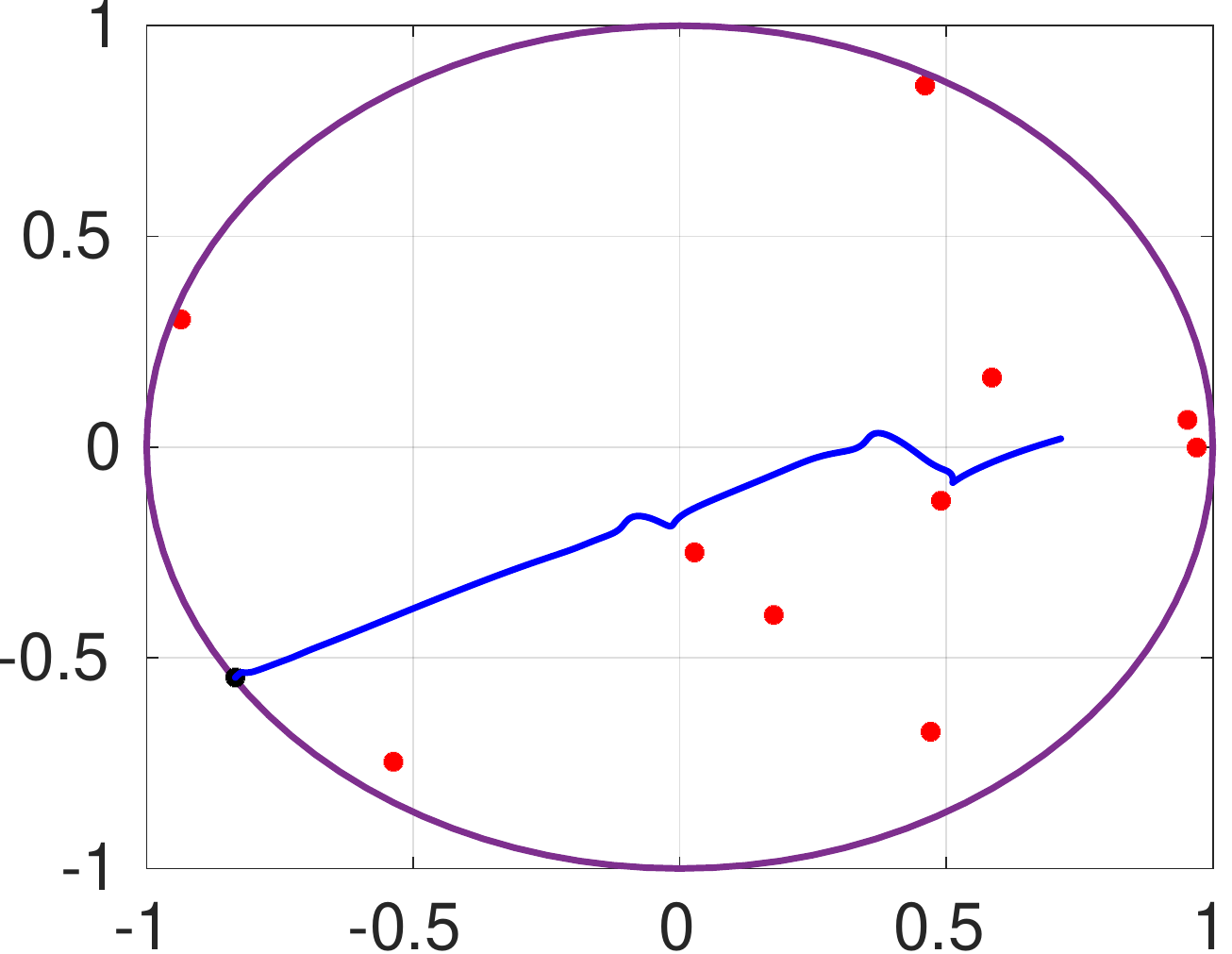}
    \caption{Transportation of object 1 from $\pi_2$ to $\pi_4$ by robots 1 and 2 in the original (left) and the transformed point-world (right) environment. In the original environment, the  circular volume of the coupled object-robots system has been transferred to the obstacles and workspace boundary.}
    \label{fig:transportation_ex}
\end{figure}

\begin{figure}
    \centering
    \includegraphics[width=.255\textwidth]{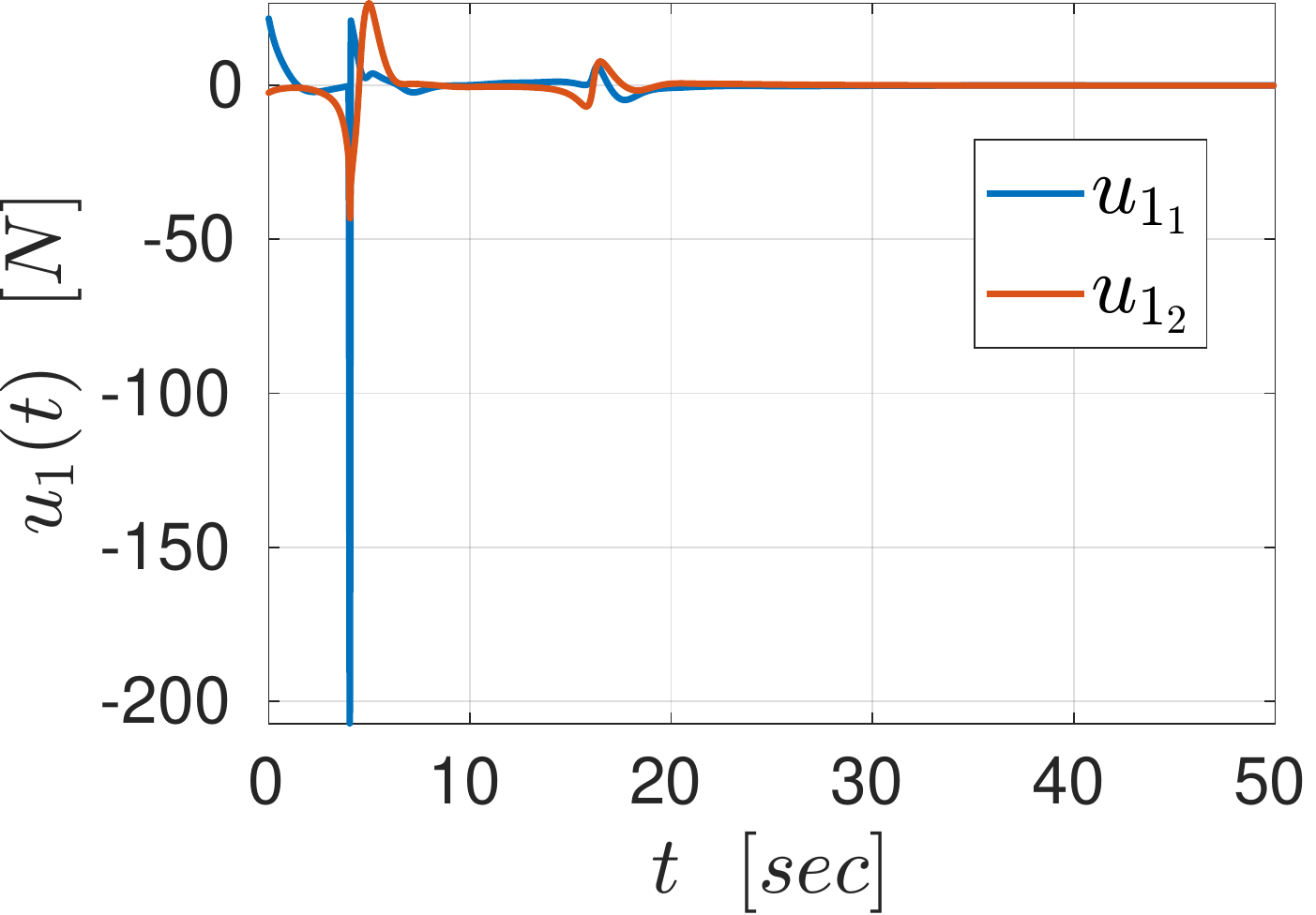}
    \includegraphics[width=.225\textwidth]{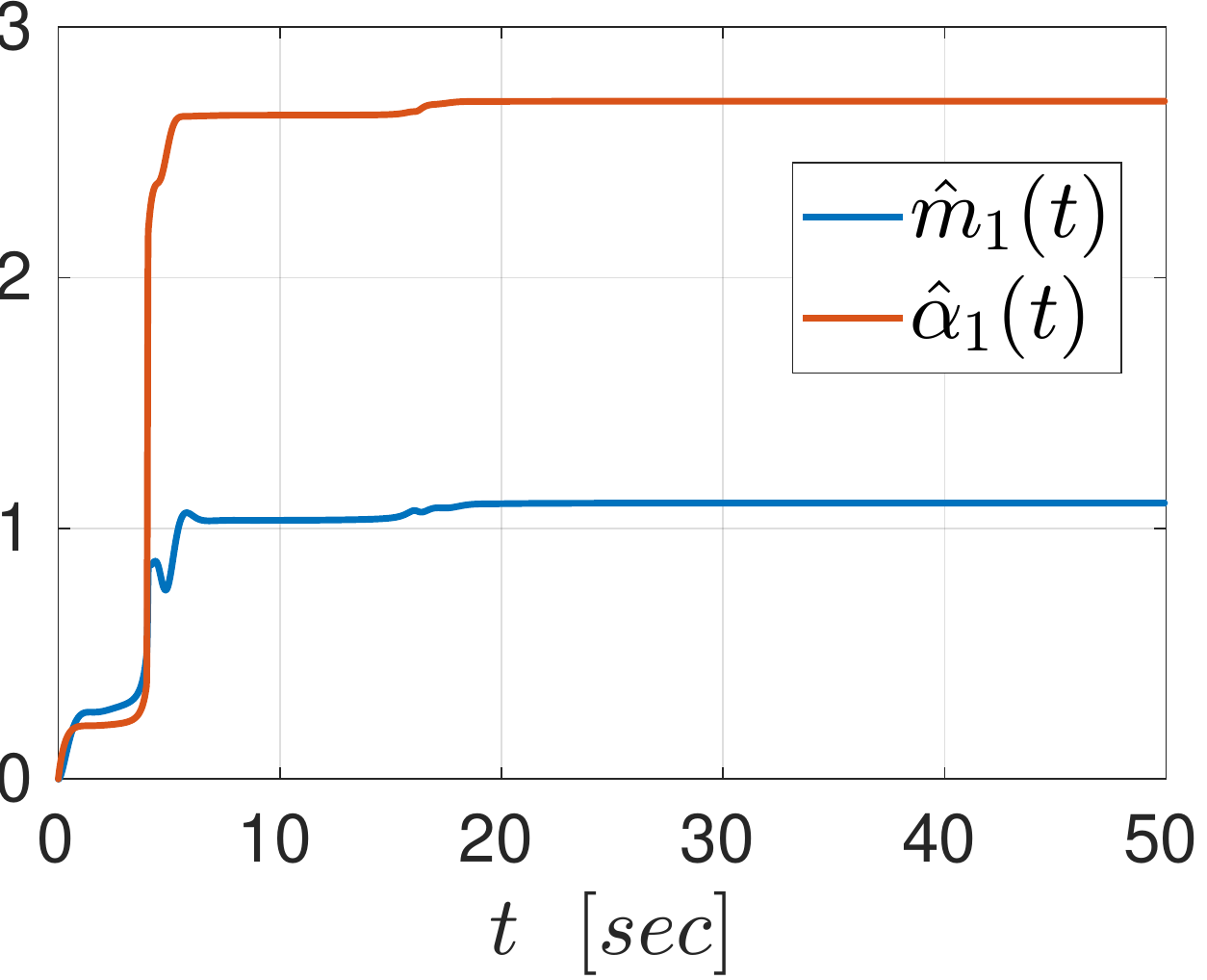}
    \caption{The evolution of the control inputs $u_1(t)$ (left) and adaptation signals $\hat{m}_1(t)$, $\hat{\alpha}_1(t)$ (right) of robot 1 in the cooperative object transportation.}
    \label{fig:transportation_signals}
\end{figure}

The resulting transition system consists of $3,112$ reachable states and $154,960$ transitions and it was created within $1.71$ minutes. We set the following formula for the multi-robot-object system:  
%The robots are responsible for accomplishing the following LTL task:\footnote{\textcolor{red}{[I know that I am not using the correct notation here, but we can fix it later :) ]}}
\begin{equation}\label{eq:task1}
    \phi=\square\Diamond \bar{\psi}_1 \wedge \square\Diamond \bar{\psi}_2 \wedge \square\Diamond \bar{\psi}_2 \wedge \Diamond \bar{\psi}_4 \wedge (\neg \bar{\psi}_4 \mathcal{U} \bar{\psi}_5 )
    %\phi=\square\Diamond \pi_{21195} \wedge \square\Diamond \pi_{24011} \square\Diamond \pi_{7542} \wedge (\neg \pi_{21195} \mathcal{U} \pi_{7542})
\end{equation}
with 
\begin{itemize}
    \item $\bar{\psi}_1 = \{ ``1\text{-}\pi_1",``2\text{-}\pi_1", ``3\text{-}\pi_1", ``O_1\text{-}\pi_1", ``O_2\text{-}\pi_4"  \}$
    \item $\bar{\psi}_2 = \{ ``1\text{-}\pi_1",``2\text{-}\pi_3", ``3\text{-}\pi_1", ``O_1\text{-}\pi_1", ``O_2\text{-}\pi_3"  \}$
    \item $\bar{\psi}_3 = \{ ``1\text{-}\pi_3",``2\text{-}\pi_3", ``3\text{-}\pi_3", ``O_1\text{-}\pi_4", ``O_2\text{-}\pi_3"  \}$
    \item $\bar{\psi}_4 = \{ ``1\text{-}\pi_2",``2\text{-}\pi_4", ``3\text{-}\pi_1", ``O_1\text{-}\pi_4", ``O_2\text{-}\pi_2"  \}$
    \item $\bar{\psi}_5 = \{ ``1\text{-}\pi_4",``2\text{-}\pi_4", ``3\text{-}\pi_2", ``O_1\text{-}\pi_3", ``O_2\text{-}\pi_4"  \}$
\end{itemize}
%where $\pi_i$ is true at state $i\in\{1,\dots,24012\}$ of the product TS. Specifically, (i) $\pi_{21195}$ is true when the state of the robots and objects is $\bar{\pi}_s=(3, 2, 3, 2)$, and $\bar{\pi}_s=(2, 2, 2)$, respectively, and the grasping variables are $(0, 3, 0, 3)$; (ii) $\pi_{24011}$ is true when the state of the robots and objects is $\bar{\pi}_s=(3, 3, 3, 3)$, and $\bar{\pi}_s=(3, 3, 3)$, respectively, and the grasping variables are $ (3, 3, 3, 3)$; and (iii) $\pi_{7542}$ is true when the state of the robots and objects is $\bar{\pi}_s=(1, 3, 3, 2)$, and $\bar{\pi}_s=(1, 2, 3)$, respectively, and the grasping variables are $(1, 3, 3, 0)$. 
In words, the mission specification in \eqref{eq:task1} requires the robots and objects to satisfy $\bar{\psi}_1$, $\bar{\psi}_2$, and $\bar{\psi}_3$ infinitely often, satisfy $\bar{\psi}_4$ eventually, and satisfy $\bar{\psi}_5$ before satisfying $\bar{\psi}_4$. 
%infinitely often but $\pi_{21195}$ should be satisfied for the first time only after $\pi_{7542}$ is satisfied. 
The LTL formula in \eqref{eq:task1} corresponds to a NBA with $6$ states - among which one is a final state - and $18$ transitions. STyLuS* found the first feasible prefix and suffix path within $1.23$ minutes and $0.64$ minutes on average. 
The action path of the robots is depicted in Table \ref{table:Path 1}, starting from $\pi_{s,1}$ and satisfying $\{ ``1\text{-}\pi_1", ``2\text{-}\pi_2", ``3\text{-}\pi_4", ``O_1\text{-}\pi_2", ``O_2\text{-}\pi_1" \}$. 

We further illustrate the continuous control design. In particular, we consider the actions of robot navigation and object transportation. Firstly, we examine the navigation of robot 1 from $\pi_1$ to $\pi_3$. The results are depicted in Figs. \ref{fig:navigation_ex}, \ref{fig:navigation_signals}. The left part of Fig. \ref{fig:navigation_ex} shows 
the trajectory of robot $1$ in the environment, where the obstacles and boundary have absorbed the the spherical volume of the robot; the right part of Fig. \ref{fig:navigation_ex} shows 
the trajectory of robot $1$ in the transformed point world, where
the obstacles are represented by points. Fig. \ref{fig:navigation_signals} depicts the control input $u_1(t)$ (left) and the evolution of the adaptation signals $\hat{m}_1(t)$, $\hat{\alpha}_1(t)$ (right). 

We further examine the transportation of object $1$ by robots $1$ and $2$ from $\pi_2$ to $\pi_4$, where we chose the load-sharing coefficients $\mathsf{cf}_1 = \mathsf{cf}_2 = 0.5$. The results are depicted in Figs. \ref{fig:transportation_ex}, \ref{fig:transportation_signals}. The left part of Fig. \ref{fig:transportation_ex} shows 
the trajectory of the coupled object-robots system  in the environment, where the obstacles and boundary have absorbed the its spherical volume; the right part of Fig. \ref{fig:navigation_ex} shows 
the trajectory of the coupled system in the transformed point world, where
the obstacles are represented by points. Fig. \ref{fig:navigation_signals} depicts the control input $u_1(t)$ (left) and the evolution of the adaptation signals $\hat{m}_1(t)$, $\hat{\alpha}_1(t)$ (right) of the robot 1, which are identical to the ones of robot 2.

%\footnote{\textcolor{red}{I have stored these paths as sequence of TS states in case we need them.}}

\begin{table}[t]
	\caption{The agent actions for the discrete path of the first case study}
	\label{table:Path 1}
	\centering
	\begin{tabular}{|p{0.45cm}|p{3.15cm}|p{0.45cm}|p{3.15cm}|} 
		%\hline
		%\textit{State} & \textit{Agent} $1$ \textit{region} & \textit{Agent} $2$ \textit{region} & \textit{Object} $1$ \textit{region} &  \textit{Object} $2$ \textit{region} & \textit{Agent} $1$ \textit{grasping} & \textit{Agent} $2$ \textit{grasping} & \textit{Atomic} \textit{propositions}  \\ [0.5ex]
		\hline
		 $\pi_{s,\ell} $ & Actions & $\pi_{s,\ell} $ & Actions \\ [0.5ex] 
		\hline\hline 		
		$\pi_{s,1}$ & $-$ & $\pi_{s,35}$ & $\pi_2 \to_{1} \pi_1$, $2 \xrightarrow{r} 2$, $3 \xrightarrow{r} 2$  \\ [0.5ex] 
		\hline
		$\pi_{s,2}$ & $1 \xrightarrow{g} 2$, $\pi_4 \to_3 \pi_1$  & $\pi_{s,36}$ & $1 \xrightarrow{g} 1$, $\pi_3 \to_3 1$   \\ [0.5ex] 
		\hline
		$\pi_{s,3}$ & $\pi_3 \to_{2} \pi_2$ $3 \xrightarrow{g} 2$   & $\pi_{s,37}$ & $\pi_3 \to_2 \pi_2$, $3 \xrightarrow{g} 1$    \\ [0.5ex]
		\hline
		$\pi_{s,4}$ &  $\pi_1 \xrightarrow{T}_{\{1,3\},2} \pi_4$, $2 \xrightarrow{g} 1$  &  $\pi_{s,38}$ & $\pi_1 \xrightarrow{T}_{\{1,3\},1} \pi_4$  \\ [0.5ex]
		\hline
		$\pi_{s,5}$ & $1 \xrightarrow{r} 2$, $2 \xrightarrow{r} 1$  & $\pi_{s,39}$ & $\pi_2 \to_2 \pi_3$, $1 \xrightarrow{r} 1$, $3 \xrightarrow{r} 1$
		  \\ [0.5ex]
		\hline
		$\pi_{s,6}$ & $\pi_4 \rightarrow_1 \pi_2$, $2 \xrightarrow{g} 1$, $3 \xrightarrow{r} 2$  & $\pi_{s,40}$ &$\pi_4 \to_3 \pi_3$   \\ [0.5ex]
		\hline
		$\pi_{s,7}$ & $1 \xrightarrow{g} 1$   & $\pi_{s,41}$ &$\pi_4 \to_1 \pi_3$, $3 \xrightarrow{g} 2$ \\ [0.5ex]
		\hline 
		$\pi_{s,8}$ & $\pi_2 \xrightarrow{T}_{\{1,2\},1} \pi_3$ & $\pi^\star_{s,42}$ & $ \pi_3 \to_2 \pi_4$, $3 \xrightarrow{r} 2$  \\ [0.5ex]
		\hline 
		$\pi_{s,9}$ & $1 \xrightarrow{r} 1$, $\pi_4 \to_{3} \pi_2$   & $\pi^\star_{s,43}$ & $ \pi_3 \to_3 \pi_4$  \\ [0.5ex]
		\hline
		$\pi_{s,10}$ & $\pi_3 \to_{1} \pi_4$, $2 \xrightarrow{r} 1$  & $\pi^\star_{s,44}$ & $2 \xrightarrow{g} 1$, $3 \xrightarrow{g} 1$  \\ [0.5ex]
		\hline 
		$\pi_{s,11}$ & $\pi_3 \to_{2} \pi_4$ & $\pi^\star_{s,45}$ & $\pi_4 \xrightarrow{T}_{\{2,3\},1} \pi_1$  \\ [0.5ex]
		\hline
		$\pi_{s,12}$ & $1 \xrightarrow{g} 2$, $\pi_2 \to_{3} \pi_4$  & 	$\pi^\star_{s,46}$ & $2 \xrightarrow{r} 1$, $2 \xrightarrow{r} 3$   \\ [0.5ex]	
		\hline 
		$\pi_{s,13}$ & $\pi_4 \to_{2} \pi_3 $, $3 \xrightarrow{g} 2$ & $\pi^\star_{s,47}$ & $ \pi_1 \to_3 \pi_3$  \\ [0.5ex]
		\hline
		$\pi_{s,14}$ & $\pi_4 \xrightarrow{T}_{\{1,3\},2} \pi_2$ & $\pi^\star_{s,48}$ &  $1 \xrightarrow{g} 2$,  $2 \xrightarrow{g} 1$,  $3 \xrightarrow{g} 2$   \\ [0.5ex]
		\hline
		$\pi_{s,15}$ & $1 \xrightarrow{r} 2$, $2 \xrightarrow{g} 1$, $3 \xrightarrow{r} 2$ &$\pi^\star_{s,49}$ & $\pi_3 \xrightarrow{T}_{\{1,3\},2} \pi_4$    \\ [0.5ex]
		\hline
		$\pi_{s,16}$ &  $2 \xrightarrow{r} 1$, $\pi_2 \to_{3} \pi_3 $ & $\pi^\star_{s,50}$ & $3 \xrightarrow{r} 2$  \\ [0.5ex]
		\hline
		$\pi_{s,17}$ &  $2 \xrightarrow{g} 1$, $3 \xrightarrow{g} 1$ & $\pi^\star_{s,51}$ & $1 \xrightarrow{r} 2$, $2 \xrightarrow{r} 1$, $\pi_4 \to_3 \pi_1$ \\ [0.5ex]
		\hline
		$\pi_{s,18}$ &  $\pi_3 \xrightarrow{T}_{\{2,3\},1} \pi_4$ & $\pi^\star_{s,52}$  &  $\pi_4 \to_1 \pi_1$, $3 \xrightarrow{g} 1$  \\ [0.5ex]
		\hline
		$\pi_{s,19}$ &  $2 \xrightarrow{r} 1$, $3 \xrightarrow{r} 1$  & $\pi^\star_{s,53}$ & $\pi_1 \to_2 \pi_4$, $3 \xrightarrow{r} 1$   \\ [0.5ex]
		\hline
		$\pi_{s,20}$ &  $1 \xrightarrow{g} 2$, $2 \xrightarrow{g} 1$, $\pi_4 \to_{3} \pi_1$  & $\pi^\star_{s,54}$ & $\pi_1 \to_3 \pi_4$ \\ [0.5ex] 
		\hline
		$\pi_{s,21}$ & $\pi_1 \to_{3} \pi_4$  & $\pi^\star_{s,55}$ & $1 \xrightarrow{g} 1$, $2 \xrightarrow{g} 2$, $3 \xrightarrow{g} 2$ \\ [0.5ex] 
		\hline
		$\pi_{s,22}$ & $3 \xrightarrow{g} 1$  & $\pi^\star_{s,56}$ & $1 \xrightarrow{r} 1$, $\pi_4 \xrightarrow{T}_{\{2,3\},2} \pi_3$   \\ [0.5ex] 
		\hline
		$\pi_{s,23}$ & $1 \xrightarrow{r} 2$, $\pi_4 \xrightarrow{T}_{\{2,3\},1} \pi_1$  & $\pi^\star_{s,57}$ & $2 \xrightarrow{r} 2$, $3 \xrightarrow{r} 2$   \\ [0.5ex] 
		\hline
		$\pi_{s,24}$ & $2 \xrightarrow{r} 1$, $3 \xrightarrow{r} 1$  & $\pi^\star_{s,58}$  &  $1 \xrightarrow{g} 1$, $\pi_3 \to_{3} \pi_1$  \\ [0.5ex] 
		\hline
		$\pi_{s,25}$ & $1 \xrightarrow{g} 2$, $\pi_1 \to_{3} \pi_2$  & $\pi^\star_{s,59}$ & $\pi_3 \to_{2} \pi_2$, $3 \xrightarrow{g} 1$   \\ [0.5ex] 
		\hline
		$\pi_{s,26}$ & $1 \xrightarrow{g} 2$, $\pi_1 \to_{2} \pi_3$, $3 \xrightarrow{g} 2$ &  $\pi^\star_{s,60}$ & $\pi_1 \xrightarrow{T}_{\{1,3\},1} \pi_4$  \\ [0.5ex] 
		\hline
		$\pi_{s,27}$ & $\pi_2 \xrightarrow{T}_{\{1,3\},2} \pi_4$ & $\pi^\star_{s,61}$ & $1 \xrightarrow{r} 1$, $\pi_2 \to_{2} \pi_3$, $3 \xrightarrow{r} 1$  \\ [0.5ex] 
		\hline
		$\pi_{s,28}$ & $1 \xrightarrow{r} 2$, $\pi_3 \to_{2} \pi_1$, $3 \xrightarrow{r} 2$ & $\pi^\star_{s,62}$ & $\pi_4 \to_{3} \pi_3$  \\ [0.5ex] 
		\hline
		$\pi_{s,29}$ & $\pi_4 \to_{3} \pi_1$ &  $\pi^\star_{s,63 }$ & $\pi_4 \to_{1} \pi_3$, $3 \xrightarrow{g} 2$  \\ [0.5ex] 
		\hline
		$\pi_{s,30}$ & $\pi_4 \to_{1} \pi_1$, $3 \xrightarrow{g} 1$ & &  \\ [0.5ex] 
		\hline
		$\pi_{s,31}$ & $\pi_1 \to_{2} \pi_4$, $3 \xrightarrow{r} 1$  & &  \\ [0.5ex] 
		\hline
		$\pi_{s,32}$ & $\pi_1 \to_{3} \pi_4$  & &  \\ [0.5ex] 
		\hline
		$\pi_{s,33}$ & $\pi_1 \to_1 \pi_2$, $2 \xrightarrow{g} 2$,$3 \xrightarrow{g} 2$    & &  \\ [0.5ex] 
		\hline
		$\pi_{s,34}$ & $\pi_4 \xrightarrow{T}_{\{2,3\},2} \pi_3$ & &  \\ [0.5ex] 
		\hline
	\end{tabular}
\end{table}

\subsection{Case Study II:} 

In this case study, we consider $N=4$ robots, $K=3$ regions and $M=3$ objects. The regions of interest are circular centered at $(88,-280)$m, $(100,-160)$m, and $(200, -130)$m, with radius equal to $4$m, as in the previous case study. 
The robots' and object' mass is taken as $1$kg and $0.25$kg, respectively, while their spherical volume's radii as $0.75$m and $0.2$m, respectively. The power capabilities of the robots are $2, 3, 4, 4$, respectively, and the power required for each object is $5, 8, 10.5$, respectively.
Initially, the robots are located in regions $\pi_{init(1)} = \pi_1$, $\pi_{init(2)}=\pi_3$, $\pi_{init(3)}=\pi_2$, and $\pi_{init(4)}=\pi_2$, respectively, whereas the objects are located in regions $\pi_{init_o(1)}=\pi_2$, $\pi_{init_o(2)}=\pi_1$, and $\pi_{init_o(3)}=\pi_3$, respectively.

The resulting transition system consists of $24,012$  reachable states and $1,805,202$ transitions and it was created within 1.52 hours.
We set the following formula for the multi-robot-object system:  
%The robots are responsible for accomplishing the following LTL task:\footnote{\textcolor{red}{[I know that I am not using the correct notation here, but we can fix it later :) ]}}
%\phi= & \square\Diamond \big( \bar{\psi}_1 \land \Diamond ( \bar{\psi}_1 \lor \bar{\psi}_2 ) \big) \land \square\Diamond (\bar{\psi}_1) \land  \square\Diamond (\bar{\psi}_3) \land \square\Diamond (\bar{\psi}_4) \notag \\ & \land ( \neg \bar{\psi}_1 \mathcal{U} \bar{\psi}_5 ) \land \square (\neg \bar{\psi}_6) \land \Diamond ( \bar{\psi}_7 \lor \bar{\psi}_8  ) \land \square \Diamond \bar{\psi}_5
\begin{align} \label{eq:task2}
    \phi= & \square\Diamond (\bar{\psi}_1 \lor \bar{\psi}_2) \land  \square\Diamond (\bar{\psi}_3) \land \square\Diamond (\bar{\psi}_4) \notag\land ( \neg \bar{\psi}_1 \mathcal{U} \bar{\psi}_5 )  \\ 
    &  \land \square (\neg \bar{\psi}_6) \land \Diamond ( \bar{\psi}_7 \lor \bar{\psi}_8  ) \land \square \Diamond \bar{\psi}_5
    %\wedge \square\Diamond \bar{\psi}_2 \square\Diamond \bar{\psi}_2 \wedge \Diamond \bar{\psi}_4  (\neg \bar{\psi}_4 \mathcal{U} \bar{\psi}_5 )
    %\phi=\square\Diamond \pi_{21195} \wedge \square\Diamond \pi_{24011} \square\Diamond \pi_{7542} \wedge (\neg \pi_{21195} \mathcal{U} \pi_{7542})
\end{align}
with 
\begin{itemize}
    \item   
    $\begin{aligned}[t]
    \bar{\psi}_1 = \{\hspace{.25cm} &``1\text{-}\pi_1",``2\text{-}\pi_1", ``3\text{-}\pi_1", ``4\text{-}\pi_1", \\ 
    &``O_1\text{-}\pi_1", ``O_2\text{-}\pi_1", ``O_3\text{-}\pi_1" \hspace{.6cm}  \}
    \end{aligned}$
    \item 
    $\begin{aligned}[t]
    \bar{\psi}_2 = \{\hspace{.25cm} &``1\text{-}\pi_1",``2\text{-}\pi_4", ``3\text{-}\pi_3", ``4\text{-}\pi_2", \\ 
    &``O_1\text{-}\pi_1", ``O_2\text{-}\pi_1", ``O_3\text{-}\pi_2" \hspace{.6cm}  \}
    \end{aligned}$
    \item
    $\begin{aligned}[t]
    \bar{\psi}_3 = \{\hspace{.25cm} &``1\text{-}\pi_2",``2\text{-}\pi_2", ``3\text{-}\pi_2", ``4\text{-}\pi_3", \\ 
    &``O_1\text{-}\pi_1", ``O_2\text{-}\pi_1", ``O_3\text{-}\pi_3" \hspace{.6cm}  \}
    \end{aligned}$
    \item
    $\begin{aligned}[t]
    \bar{\psi}_4 = \{\hspace{.25cm} &``1\text{-}\pi_1",``2\text{-}\pi_1", ``3\text{-}\pi_2", ``4\text{-}\pi_3", \\ 
    &``O_1\text{-}\pi_3", ``O_2\text{-}\pi_3", ``O_3\text{-}\pi_1" \hspace{.6cm}  \}
    \end{aligned}$
    \item
    $\begin{aligned}[t]
    \bar{\psi}_5 = \{\hspace{.25cm} &``1\text{-}\pi_2",``2\text{-}\pi_2", ``3\text{-}\pi_3", ``4\text{-}\pi_3", \\ 
    &``O_1\text{-}\pi_2", ``O_2\text{-}\pi_3", ``O_3\text{-}\pi_3" \hspace{.6cm}  \}
    \end{aligned}$
    \item
    $\begin{aligned}[t]
    \bar{\psi}_6 = \{\hspace{.25cm} &``1\text{-}\pi_2",``2\text{-}\pi_2", ``3\text{-}\pi_2", ``4\text{-}\pi_2", \\ 
    &``O_1\text{-}\pi_3", ``O_2\text{-}\pi_3", ``O_3\text{-}\pi_3" \hspace{.6cm}  \}
    \end{aligned}$
    \item
    $\begin{aligned}[t]
    \bar{\psi}_7 = \{\hspace{.25cm} &``1\text{-}\pi_3",``2\text{-}\pi_3", ``3\text{-}\pi_2", ``4\text{-}\pi_1", \\ 
    &``O_1\text{-}\pi_3", ``O_2\text{-}\pi_1", ``O_3\text{-}\pi_2" \hspace{.6cm}  \}
    \end{aligned}$
    \item
    $\begin{aligned}[t]
    \bar{\psi}_8 = \{\hspace{.25cm} &``1\text{-}\pi_4",``2\text{-}\pi_4", ``3\text{-}\pi_3", ``4\text{-}\pi_2", \\ 
    &``O_1\text{-}\pi_3", ``O_2\text{-}\pi_3", ``O_3\text{-}\pi_3" \hspace{.6cm}  \}
    \end{aligned}$
    \end{itemize}

The LTL formula in \eqref{eq:task2} corresponds to a NBA with $8$ states - among which one is a final state - and $27$ transitions. STyLuS* found the first feasible prefix and suffix path within $3.73$ minutes and $1.43$ minutes on average. We omit the action path of the robots for ease of exposition. 

%In this case study, the initial setup is the same as in the previous case study. In this scenario, the mission for the robots is captured by the following LTL formula:
%\begin{align}\label{eq:task2}
 %   \phi=&\square\Diamond \pi_{21195} \wedge \square\Diamond \pi_{24011} \square\Diamond \pi_{7542} \wedge \square\Diamond(\pi_{9631} \wedge \Diamond (\pi_{19705})) \nonumber\\&\wedge
 %   (\neg \pi_{24011} \mathcal{U} \pi_{19705}),
%\end{align}
%where (i) $\pi_{9631}$ is true when the state of the robots and objects is $\bar{\pi}_s=(2, 1, 2, 3)$, and $\bar{\pi}_s=(2, 1, 2)$, respectively, and the grasping variables are $(1, 0, 1, 0)$ and (ii) $\pi_{19705}$ is true when the state of the robots and objects is $\bar{\pi}_s=(3, 2, 1, 3)$, and $\bar{\pi}_s=(3, 1, 2)$, respectively, and the grasping variables are $(0, 0, 0, 1)$.  In words, the mission specification in \eqref{eq:task1} requires the robots and objects to (i) satisfy $\pi_{21195}$, $\pi_{24011}$, and $\pi_{7542}$ in any order infinitely often, (ii) avoid $\pi_{24011}$ until $\pi_{19705}$ is satisfied, and (iii) satisfy $\pi_{9631}$ and $\pi_{19705}$ in this order infinitely often.  The LTL formula in \eqref{eq:task2} corresponds to a NBA with $12$ states - among which one is a final state - and $15$ transitions. STyLuS* found the first feasible prefix and suffix path within $2.61$ minutes and $4.01$ minutes on average.
%\subsection{Case Study III:} In this case study, we consider $N=5$ robots and $K=3$ and $M=3$. [\textcolor{red}{currently running...}]

\section{Conclusion} \label{sec:conclusion}

We propose an algorithm for the control and planning of multi-robot-object systems subject LTL tasks. We develop adaptive feedback-control protocols for robot navigation and cooperative object transportation, which enable the abstraction of the underlying continuous dynamics to a finite transition system. We compose the transition system with an automaton that represents the LTL task and 
use a sampling-based planner to derive an optimal task-satisfying plan for the robots.  

\bibliographystyle{IEEEtran}
\bibliography{references}
\end{document}